\documentclass[a4paper,11pt]{article}
\pdfoutput=1
\usepackage{jheppub}
\usepackage[T1]{fontenc}
\usepackage[english]{babel}
\usepackage{amsthm,amsfonts}
\usepackage{dsfont}
\hyphenchar\font=\string"7F
\usepackage{tensor}
\usepackage[mathstyleoff]{breqn}
\usepackage{slashed}
\usepackage{stmaryrd}
\usepackage{enumerate}
\usepackage[normalem]{ulem}
\usepackage{tikz}
\usepackage[mathscr]{euscript}
\usepackage{mathrsfs}
\usepackage{todonotes}

\DeclareMathAlphabet{\mathdsl}{U}{bbm}{m}{sl}

\newcommand{\dd}{\mathrm{d}}

\renewcommand{\sc}{\underline{c}}

\makeatletter
\DeclareFontFamily{OMX}{MnSymbolE}{}
\DeclareSymbolFont{MnLargeSymbols}{OMX}{MnSymbolE}{m}{n}
\SetSymbolFont{MnLargeSymbols}{bold}{OMX}{MnSymbolE}{b}{n}
\DeclareFontShape{OMX}{MnSymbolE}{m}{n}{
    <-6>  MnSymbolE5
   <6-7>  MnSymbolE6
   <7-8>  MnSymbolE7
   <8-9>  MnSymbolE8
   <9-10> MnSymbolE9
  <10-12> MnSymbolE10
  <12->   MnSymbolE12
}{}
\DeclareFontShape{OMX}{MnSymbolE}{b}{n}{
    <-6>  MnSymbolE-Bold5
   <6-7>  MnSymbolE-Bold6
   <7-8>  MnSymbolE-Bold7
   <8-9>  MnSymbolE-Bold8
   <9-10> MnSymbolE-Bold9
  <10-12> MnSymbolE-Bold10
  <12->   MnSymbolE-Bold12
}{}

\let\llangle\@undefined
\let\rrangle\@undefined
\DeclareMathDelimiter{\llangle}{\mathopen}%
                     {MnLargeSymbols}{'164}{MnLargeSymbols}{'164}
\DeclareMathDelimiter{\rrangle}{\mathclose}%
                     {MnLargeSymbols}{'171}{MnLargeSymbols}{'171}
\makeatother

\allowdisplaybreaks
\newcommand{\HIDDEN}[1]{}

\title{\boldmath Generalised Cosets}
\preprint{MPP-2019-251}
 
\author[a,b,c]{Saskia Demulder,}
\author[d]{Falk Hassler,}
\author[a]{Giacomo Piccinini,}
\author[a,b]{and Daniel C. Thompson}

\emailAdd{sademuld@mpp.mpg.de}
\emailAdd{falk@fhassler.de}
\emailAdd{g.piccinini.987589@swansea.ac.uk}
\emailAdd{D.C.Thompson@Swansea.ac.uk}

\affiliation[a]{Department of Physics, Swansea University, Swansea, SA2 8PP, U.K.}
\affiliation[b]{
Theoretische Natuurkunde, Vrije Universiteit Brussel \& The International Solvay Institutes,\\ B-1050 Brussels, Belgium}
\affiliation[c]{Max-Planck-Institut f\"ur Physik, F\"ohringer Ring 6, 80805 M\"unchen, Germany}
\affiliation[d]{George P. \& Cynthia Woods Mitchell Institute for Fundamental Physics and Astronomy,\\ Texas A\&M University, College Station, TX 77843, USA}

\abstract{Recent work has shown that two-dimensional non-linear $\sigma$-models on group manifolds with Poisson-Lie symmetry can be understood within generalised geometry as exemplars of generalised parallelisable spaces. Here we extend this idea to target spaces constructed as double cosets $M=\widetilde{G} \backslash \mathdsl{D} / H$. Mirroring conventional coset geometries, we show that on $M$ one can construct a generalised frame field and a $H$-valued generalised spin connection that together furnish an algebra under the generalised Lie derivative. This results naturally in a generalised covariant derivative with a (covariantly) constant generalised intrinsic torsion, lending itself to the construction of consistent truncations of 10-dimensional supergravity compactified on $M$. An important feature is that $M$ can admit distinguished points, around which the generalised tangent bundle should be augmented by localised vector multiplets. We illustrate these ideas with explicit examples of two-dimensional parafermionic theories and NS5-branes on a circle.}

\begin{document} 
\maketitle 
\section{Introduction}
Lower dimensional gauged supergravities (SUGRAs) are  powerful tools to address properties of holographic  quantum field theories; however,  their usage requires some care. In a typical Kaluza-Klein scenario one justifies a lower dimensional theory by dispensing modes suppressed by the scale of the internal manifold. In the context of holography, the scale of the $d$-dimensional internal manifold $M$ and of the anti-de Sitter external space are comparable and such an argument cannot be readily made. What is then required is that the lower dimensional theory retains a ``consistent truncation'' of modes from the 10- or 11-dimensional maximal SUGRA. Finding such truncation and the corresponding, in general highly, non-linear Ans\"{a}tze for the bosonic fields is involved. Recently, generalised/exceptional geometry  \cite{Hitchin:2004ut,GualtieriThesis}/\cite{Hull:2007zu,Pacheco:2008ps,Coimbra:2012af} and the closely related double/exceptional \cite{Hull:2007zu,Hohm:2010pp}/\cite{Berman:2010is,Hohm:2013pua} field theories were employed to make the process more methodical starting with \cite{Lee:2014mla,Hohm:2014qga}. A systematic treatment of such constructions was given in  \cite{Cassani:2019vcl} exploiting idea that when $M$ is equipped with a generalised $G$-structure, defined by a set of invariant tensors and a covariantly constant singlet intrinsic torsion, a consistent truncation is obtained by expanding bosonic fields in terms of the invariant tensors.

Important examples within the context of the 10-dimensional type II supergravities are manifolds with trivial generalised structure group. These admit a globally defined frame $E_A$ on the generalised tangent bundle, locally given by\footnote{In the cases of exceptional generalised geometry the generalised tangent bundle is further augmented with higher rank differential forms.}  $TM \oplus T^\star M$, and are therefore called generalised (Leibnitz) parallelisable. The action of the $T M$ part of $E_A$ on functions gives rise to a flat covariant derivative\footnote{$D_A$ does not have a spin connection, but it has an affine connection which is fixed by imposing $D_A E_B = 0$.}, $D_A$, with (covariantly) constant generalised torsion $\mathcal{F}_{AB}{}^C$ governed by the relation
\begin{equation}\label{eqn:framealge_intro}
  \mathscr{L}_{E_A} E_B = \mathcal{F}_{AB}{}^C E_C\,.
\end{equation}
Here $\mathscr{L}$ denotes the generalised Lie derivative and because the frame fields $E_A$ are invariant, all bosonic fields expanded in terms of them form a consistent truncation. The resulting lower dimensional SUGRA is maximally supersymmetric due to the trivial generalised structure group of $M$. As such truncations are conceptually closely related to Scherk-Schwarz reductions on group manifolds \cite{Scherk:1978ta,Scherk:1979zr}, they are called generalised Scherk-Schwarz reductions \cite{Grana:2012rr,Geissbuhler:2011mx,Aldazabal:2011nj}. Most notably, the torsion $\mathcal{F}_{AB}{}^C$ is in one-to-one correspondence with the embedding tensor which fixes the gauge group of the lower dimensional maximal SUGRA. The remaining challenge is to find at least one tuple ($M$, $E_A$) such that \eqref{eqn:framealge_intro} holds for a given constant generalised torsion $\mathcal{F}_{AB}{}^C$. The manifold $M$ is necessarily a coset \cite{Grana:2008yw,Lee:2014mla} but still constructing generalised parallelisable spaces is challenging. There are few examples known \cite{Dibitetto:2012rk,Lee:2014mla,Hohm:2014qga} but a systematic construction was missing until recently and instead considered on a case by case basis. In generalised geometry a complete construction of generalised parallelisable spaces was worked out in a series of papers \cite{Demulder:2018lmj,Hassler:2019wvn} in which the right coset\footnote{We choose here to work with right cosets to minimised the number of minus signs in our conventions. The conclusions of course  can be made equally well with left cosets.} $\widetilde G\backslash\mathdsl{D}$ is identified with the internal manifold $M$, where $\mathdsl{D}$ is a Lie group which admits a non-degenerate, invariant pairing of split signature for which  the subgroup $\widetilde G \subset \mathdsl{D} $ is  maximally isotropic. The constant generalised torsion is given by the structure coefficients of $\mathfrak{d}$, the Lie algebra of $\mathdsl{D}$. In this case, which we review in Section~\ref{sec:genpara}, $E_A$ can be constructed from a coset representative $m \in M$. Intriguingly, its construction shares many features with Poisson-Lie T-duality \cite{Klimcik:1995ux,Klimcik:1995dy} where the two groups $\mathdsl{D}$ and $\widetilde G$ appear naturally. In general there are multiple tuples ($M$, $E_A$), ($M'$, $E'_A$), \ldots ,  which solve \eqref{eqn:framealge_intro} and originate from different maximally isotropic subgroups $\widetilde{G}$, $\widetilde{G}'$, \ldots\,. The different sets of bosonic fields one constructs out of them are related by Poisson-Lie T-duality and give rise to the same lower dimensional SUGRA \cite{Hassler:2017yza}.

However, the kind of Poisson-Lie T-duality which appears in this context is not the most general one. It is rather a special case of the dressing coset construction \cite{Klimcik:1996np} with $M=\widetilde{G} \backslash \mathdsl{D} / H$ for a trivial $H$. Hence, motivated by the interplay between generalised geometry and Poisson-Lie T-duality in the example above, this paper shows that dressing cosets give rise to a class of new generalised geometries which are relevant for the construction of consistent truncations. They are named {\it generalised cosets} because of the analogy with conventional coset spaces. More explicitly, they admit a generalised connection $\nabla_A$ whose torsion and curvature are both parallel (covariantly constant). Along the line of discussion for generalised parallelisable spaces above, $\nabla_A$ is expressed in terms of a generalised frame $E_A$ but now also a non-vanishing spin connection $\Omega = \Omega^\alpha t_\alpha $ valued in the Lie algebra of $H$ (generated by $t_\alpha$) is required. The requirement that $\nabla_A$ has parallel torsion and curvature result in the modified frame algebra
\begin{equation}\label{eqn:framealge2_intro}
  \begin{aligned}
    \mathscr{L}_{E_A} E_B &= F_{AB}{}^C E_C + 2 \Omega^\alpha{}_{[A} F_{\alpha B]}{}^C E_C + F_{\alpha AB} \Omega^{\alpha} \, , \\
    \mathscr{L}_{\Omega^{\alpha}{}} E_B &=  (\Omega^{\beta\alpha} F_{\beta B C} + F^\alpha{}_{BC}) E^C + F_{\beta\gamma}{}^{\alpha}  \Omega^{\beta}{}_B \Omega^\gamma \, , 
  \end{aligned}
\end{equation} 
where $\Omega^\alpha = \Omega^{\alpha}{}_{A} E^A$. We discuss this algebra in Section~\ref{sec:gencoset}. Here we just emphasise that all $F$'s are constant and that the generalised structure group of a generalised coset is $H$. In addition to deriving all properties relevant for consistent truncations based on this new class of generalised geometries, we also present an explicit construction of the frame $E_A$, the connection $\Omega^{\alpha}$ and $\Omega^{\alpha\beta}$ starting from the generalised frame field on the coset $\widetilde{G}\backslash\mathdsl{D}$. A remarkable feature of this class is that it also includes examples of singular geometries. This feature happens when the action of $H$ on $\widetilde{G}\backslash\mathdsl{D}$ has fixed points, and in this scenario one can think of the construction as defining a {\em generalised orbifold}. There can also be further distinguished points, which do not have to coincide with the fixed points of the $H$ action, where the generalised tangent bundle must be enhanced by localised vector multiplets $\mathscr{A}$ with a non-vanishing field strength $\mathscr{F}$. In turn this can constitute a magnetic source from NS5-branes, visible as the failure of closure in the NS three-form flux schematically given by 
\begin{equation}
  \dd H \propto c( \mathscr{F} \overset{\wedge}{,} \mathscr{F})  \, , 
\end{equation}
where the four-form on the right-hand side involves a suitable pairing in the algebra of $H$. Similar enlargements of the generalised tangent bundle have been considered in the context of e.g. heterotic models \cite{Baraglia:2013wua,Coimbra:2014qaa}, enhancement gauged symmetry at special points of moduli space \cite{Aldazabal:2015yna,Aldazabal:2017wbk} and localised vectors living on generalised orientifold planes \cite{Blair:2018lbh}. We present these phenomena for the target space of NS5-branes on a circle \cite{Sfetsos:1998xd} in Section~\ref{sec:ns5example}.  From a mathematical point of view, the natural language to situate this present discussion is that of Courant algebroids as  in \cite{Severa:2018pag} and the ideas which we present can be considered an example of a Courant algebroid reduction  \cite{bursztyn2007reduction} whose development was strongly influenced by earlier works on gauged $\sigma$-models \cite{figueroa1994equivariant,hull1989gauged,de1987new}.
 
%%%%%%%%%%%%%%%%%%%%%%%%%%%%%%%%%%%%%%%%%
\section{Cosets and generalized parallelisable spaces} \label{sec:genpara}
Let us consider a $D$-dimensional manifold $M$ that admits the action of a Lie group $\mathdsl{D}$ corresponding to a $2 D$-dimensional Lie algebra $\mathfrak{d}$ spanned by the  linearly independent generators $T_A$, $A=1,\dots, 2 D$, obeying\footnote{For $\mathdsl{D}$ compact we will take the generators to be anti-Hermitian.}  
\begin{equation}\label{eq:alg} 
   \llbracket  T_A, T_B \rrbracket = F_{AB}{}^C T_C\,.
\end{equation}
Furthermore, we assume that $\mathfrak{d}$ is equipped with an ad-invariant, non-degenerate, bi-linear symmetric pairing, $\llangle \cdot, \cdot \rrangle$, of split signature form which we define 
\begin{equation}
  \llangle T_A, T_B \rrangle = \eta_{AB}\,. 
\end{equation}
Corresponding to each generator $T_A$ of $\mathfrak{d}$ there is a vector field $k_A$ on the coset $M=\widetilde{G}\backslash \mathdsl{D}$ that under the conventional Lie bracket furnishes the same algebra as \eqref{eq:alg}, namely 
\begin{equation}\label{eqn:algconditionsvec}
  [ k_A, k_B  ] = F_{AB}{}^C k_C \,  .
\end{equation}

Let us assume the existence of a closed three-form $H$ on $M$ upon which we do not place {\em a priori} any demands of $\mathdsl{D}$ invariance. Recall that the $H$-twisted generalised Lie derivative acts on sections $U=u+ \mu$ and $V= v + \nu$ (with $u,v$ vectors and $\mu, \nu$ one-forms) of the generalised tangent bundle as  
\begin{equation}
  \mathscr{L}_{U} V = [u , v] + \left( L_u \nu - \iota_v \dd \mu\right) - \iota_u \iota_v H \, ,
\end{equation}
where $L$ is the conventional Lie derivative. 
Further recall that the pairing of two generalised vectors is given by 
\begin{equation}
\langle U ,  V\rangle =  \iota_u \nu + \iota_v \mu \,.  
\end{equation}

Generalised parallelisable spaces are those for which we can construct on $M$ a set of $O(D,D)$ valued generalised frame fields $E_A$ for which the generalised torsion of \eqref{eqn:framealge_intro} is constant and identified with the structure constants of  $\mathfrak{d}$, i.e. 
 \begin{equation}\label{eqn:framealge_sec2}
   \mathscr{L}_{E_A} E_B = F_{AB}{}^C E_C  \qquad \text{and} \qquad
   \langle E_A ,  E_B\rangle    = \llangle T_A, T_B \rrangle    \,.
\end{equation}
Since the vector part of the generalised Lie derivative matches the conventional Lie derivative, a natural ansatz for the generalised frame fields is
\begin{equation} 
  E_A = k_A + \varphi_A \, , 
\end{equation} 
with $\varphi_A$ a set of one-forms that has to be determined. It is easy to see that the algebra \eqref{eqn:framealge_sec2} is obeyed provided \eqref{eqn:algconditionsvec} holds for the vector part and, moreover, the following conditions 
\begin{equation}\label{eq:algconditions}
  \begin{aligned}
  \iota_{k_A} \varphi_B +  \iota_{k_B} \varphi_A &= \eta_{AB} \\ 
  L_{k_A} \varphi_B &= F_{AB}{}^C \varphi_C + \iota_{k_B} \vartheta_A \, , \\
  \dd\varphi_A &= \iota_{k_A} H + \vartheta_A \, , 
  \end{aligned}
\end{equation}
are satisfied from the one-form part of the generalised frame field. Here $\vartheta_A$ denotes a suitable set of two-forms that encode the failure of $\mathdsl{D}$ invariance of the three-form since $L_{k_A} H = - \dd \vartheta_A$. If  $\vartheta_A = L_{k_A} \varpi$  for some two-form $\varpi$,  then one can perform a shift $H' = H + \dd\varpi$ and $\varphi_A' = \varphi_A - \iota_{k_A} \varpi$ so that \eqref{eq:algconditions} is solved with $\vartheta_A' = 0$. It the next subsection we show that this situation can be always be achieved. Hence without loss of generality, we will choose $H$ to be invariant under $\mathdsl{D}$. In this case, it is possible to further gauge quotients of this construction as we will do in Section~\ref{sec:gencoset}. 

\subsection{The descent from \texorpdfstring{$\mathdsl{D}$}{D}} 
In general, the construction of such generalised frame fields  would seem like a formidable task;  however, when $M$ is identified with the right coset $M=\widetilde{G}\backslash\mathdsl{D}$ for $\widetilde{G}$ a maximal isotropic subgroup\footnote{An isotropic subgroup $\widetilde{G}$ is one for which the corresponding generators, $\widetilde T^a$, satisfy the condition
\begin{equation}
  \llangle  \widetilde T^a,  \widetilde T^b \rrangle = 0 \qquad  \forall \,\,  \widetilde T^a ,  \widetilde T^b \in \widetilde{\mathfrak{g}}\,.
\end{equation}
When the subgroup $\widetilde{G}$, with  Lie algebra $\widetilde{\mathfrak{g}}$, is further said to be  maximal if $\dim \widetilde{G} = D$.} of $\mathdsl{D}$, the problem becomes tractable. In this case we can proceed algorithmically to recover the results of \cite{Hassler:2019wvn} in a fashion more suitable for the extension considered in the following sections.

To fix the notation let us denote by $m$ a representative of the coset $M=\widetilde{G}\backslash\mathdsl{D}$ and write group elements of $\mathdsl{D}$ as $\mathdsl{g} = \widetilde{g} m$ with $\widetilde{g} \in \widetilde{G}$. As a group manifold, $\mathdsl{D}$ is equipped with vector fields\footnote{In what follows, we will reserve the hatted notation to objects acting on the bundle associated to the $\mathdsl D$, descending these through the (pushforward) of the coset projector we recover the objects introduced in the last section.} $\widehat k_A$ and $\widehat  v_A$  constructed as duals to the left- and right-invariant Maurer-Cartan forms namely
\begin{equation}\label{eq:duals}
  \iota_{\widehat{k}_A} \mathdsl{g}^{-1} \dd \mathdsl{g} = T_A \, , \quad 
  \iota_{\widehat{v}_A} \dd \mathdsl{g} \mathdsl{g}^{-1} = T_A\,, 
\end{equation}
and  generating respectively the right- and left- actions. 
These satisfy the  relations
 \begin{equation}
  [\widehat{k}_A, \widehat{k}_B] = F_{AB}{}^C \widehat{k}_C \, , \quad \, 
  [\widehat{v}_A, \widehat{v}_B] = - F_{AB}{}^C \widehat{v}_C  \, , \quad \, 
  [\widehat{v}_A, \widehat{k}_B] = 0\,.
\end{equation}
To descend to the coset we define a $\widetilde{\mathfrak{g}}$-valued connection $A = \dd \widetilde{g} \widetilde{g}^{-1}$,  used to rewrite the right-invariant form 
\begin{equation}
  \dd \mathdsl{g} \mathdsl{g}^{-1} = A + \tilde{g} \dd m m^{-1} \tilde{g}^{-1}\, . 
\end{equation}
The vector fields which generate right translations on the coset now arise as the restriction 
\begin{equation}\label{eq:bark2}
  \overline{k}_A  = \widehat{k}_A - \iota_{\widehat{k}_A} A_b v^b\,.
\end{equation}
To see that these vector fields are indeed  governed by the algebra \eqref{eqn:algconditionsvec}, it is sufficient to note that $A$ is flat,  $\dd A - A\wedge A =0$, and that $\iota_{\widehat{v}^a} A_b = \delta^a_b$ as well as $L_{\widehat{v}^a} A_b = F^{ac}{}_b A_c$ hold.

As a group manifold $\mathdsl{D}$ is equipped with a bi-invariant closed three-form
\begin{equation}\label{eqn:Hgroup}
  \begin{aligned}
    \widehat{H} &=  - \frac{1}{6} \llangle \dd \mathdsl{g} \mathdsl{g}^{-1} \,\overset{\wedge}{,}  \dd \mathdsl{g} \mathdsl{g}^{-1} \wedge \dd \mathdsl{g} \mathdsl{g}^{-1} \rrangle \\
    &= - \frac{1}{2} \llangle \tilde{g}^{-1} A \tilde{g} \,\overset{\wedge}{,} \dd m m^{-1} \wedge \dd m m^{-1} \rrangle - \frac{1}{2} \llangle \tilde{g}^{-1} \dd A \tilde{g} \,\overset{\wedge}{,} \dd m m^{-1} \rrangle + \overline{H} \,,
  \end{aligned}
\end{equation}
with horizontal part\footnote{A differential form $\widehat{\phi}$ is horizontal if $\iota_{\widehat{v}^a} \widehat \phi = 0$ holds.} $\overline{H}$. In particular, the closure of $\widehat{H}$  immediately implies that $\overline{H}$ is also closed.  In addition, we can construct a right-invariant two-form, 
\begin{equation}
  \widehat{\omega} = \frac12 \llangle \dd \mathdsl{g} \mathdsl{g}^{-1}\,\overset{\wedge}{,}  {\cal K} \dd \mathdsl{g} \mathdsl{g}^{-1} \rrangle = \llangle  m^{-1} \dd m  \,\overset{\wedge}{,} \mathdsl{g}^{-1} A \mathdsl{g} \rrangle + 2 \overline{\varpi}\,,
\end{equation}
with
\begin{equation}
  \overline{\varpi} = \frac14 \llangle \dd m m^{-1} \overset{\wedge}{,} \mathcal{K} \dd m m^{-1} \rrangle \,,
\end{equation}
in which ${\cal K} $ is an involution, compatible with the pairing,  \mbox{$\llangle T_A, {\cal K} T_B \rrangle = - \llangle {\cal K} T_A, T_B \rrangle$}, whose $+1$ eigenspace is identified with $\widetilde{\mathfrak{g}}$. The involution ${\cal K}$ gives rise to a para-Hermitian structure on $\mathdsl{D}$ \cite{Hassler:2019wvn}. The exterior derivative of $\omega$ precisely encodes the vertical part of the three-from on  $\mathdsl{D}$:
\begin{equation} \label{eq:Hdecomp} 
  \widehat{H}  = \overline{H} + \dd \overline{\varpi} - \frac{1}{2}  \dd \widehat \omega\,. 
\end{equation}
This has an important consequence: since both $\widehat{H}$ and $\widehat{\omega}$ are invariant under $\widehat{k}_A$ so too is the combination $\overline{H}' = \overline{H} + \dd \overline{\varpi}$; moreover, $\overline{H}$ and $\overline{\varpi}$  being horizontal,  we have that $L_{\overline{k}_A} \overline{H}' =0$. Making use of \eqref{eq:duals} and the isotropy of $\widetilde{\mathfrak{g}}$ we find the useful identity 
\begin{equation}\label{eq:identi}
  \llangle  T_A, \mathdsl{g}^{-1} A \mathdsl{g} \rrangle = \iota_{\widehat{k}_A} (\widehat{\omega} - 2 \overline{\varpi}) + \llangle  T_A, m^{-1} \dd m\rrangle - \llangle  \iota_{\widehat{k}_A} m^{-1}\dd m, m^{-1} \dd m \rrangle \, .
\end{equation} 
A further contraction of this identity returns, after exploiting again isotropy, 
\begin{equation}\label{eq:iotaAphiB}
  \iota_{\overline{k}_A} \overline{\varphi}_B = \frac12 \llangle T_A, T_B \rrangle  - \frac12 \iota_{\widehat{k}_A} \iota_{\widehat{k}_B} ( \widehat\omega - 2 \overline{\varpi})\, , 
\end{equation}
with $\overline{\varphi}_A$ given as 
\begin{equation}\label{eqn:varphiA}
  \overline{\varphi}_A = \llangle T_A, m^{-1} \dd m\rrangle - \frac12 \llangle \iota_{\overline{k}_A} \dd m m^{-1} , \dd m m^{-1} \rrangle \,.
\end{equation}
The process is algorithmic: one needs only to take as an input either $\widehat{H}$ or $\widehat{\omega}$ defined on $\mathdsl{D}$ and the rest follows. Taking the exterior derivative of eq.~\eqref{eq:identi}, and using that $\widehat{\omega}$ is invariant yields 
\begin{equation}
  -\llangle T_A,  \mathdsl{g}^{-1} \dd \mathdsl{g} \wedge \mathdsl{g}^{-1} \dd \mathdsl{g}\rrangle =  2 \dd \overline{\varphi}_{A} -  \iota_{\widehat k_A} \dd(\widehat{\omega} - 2 \overline{\varpi}) - 2 L_{\overline{k}_A} \overline{\varpi} \, , 
\end{equation}
which after comparing with  eq.~\eqref{eq:Hdecomp} implies that  $\dd \overline{\varphi}_A = \iota_{\overline{k}_A}\overline{H} + L_{\overline{k}_A} \overline{\varpi}$. Combining the above identities we eventually obtain
\begin{equation}
  L_{\overline{k}_A} \overline{\varphi}_B = F_{AB}{}^C \overline{\varphi}_C + \iota_{\overline{k}_A} L_{\overline{k}_B} \overline{\varpi} \,  .
\end{equation}

We are still working with quantities on the full group $\mathdsl{D}$, but the generalised frame fields we want to construct are defined on the coset $\widetilde{G}\backslash \mathdsl{D}$. For this coset, we have the projection $\pi: \mathdsl{D} \rightarrow \widetilde{G}\backslash \mathdsl{D}$ and the local sections $\sigma: \widetilde{G}\backslash \mathdsl{D} \rightarrow \mathdsl{D}$ which are chosen such that the pullback $\sigma^* A$ vanishes. This is always possible because $A$ is pure gauge. Equipped with these two maps, the quantities discussed at the beginning of this section are
\begin{equation}\label{eqn:ingredientsgenframe}
  k_A = \pi_* \overline{k}_A \,, \qquad
  \varphi_A = \sigma^* \overline{\varphi}_A \,, \qquad
  H = \sigma^* \overline{H} \qquad \text{and} \qquad
  \varpi = \sigma^* \overline{\varpi}\,.
\end{equation}
They satisfy the required relations eq.~\eqref{eq:alg} and eq.~\eqref{eq:algconditions} with the identification $\vartheta_A = L_{k_A} \varpi$.  In the examples we consider later $\varpi= 0$, however more generally one can use the freedom described below   eq.\eqref{eq:algconditions} to absorb $ \vartheta_A$ into a redefinition of $H$ and $\varphi_A$, and in what follows we shall assume this has been done.

\subsection{Restriction to \texorpdfstring{$\mathdsl{D}$}{D} a Drinfel'd double} 
A refinement  occurs when $\mathdsl{D}$ is a Drinfel'd double such that $\mathfrak{d} = \widetilde{\mathfrak{g}} + \mathfrak{g}$ is a decomposition into two maximally isotropic subalgebras. This is the setting of Poisson-Lie T-duality discussed in \cite{Klimcik:1995dy}.  Here  $M=\widetilde{G}/\mathdsl{D} \cong G= \exp \mathfrak{g}$ and the coset representative $m$ is identified with a group element   $g \in G$. Since the one-forms   $g^{-1}  \dd g$ are $\mathfrak{g}$-valued and since $\mathfrak{g}$ is isotropic $H$, $\varpi$ and $\vartheta_A$ all vanish. Denoting the components of the adjoint action of $g$ on $\mathfrak{d}$, in a basis where $\llangle T_a , \widetilde{T}^b \rrangle = \delta_{a}{}^b$, by $M_A{}^B T_B = g T_A g^{-1}$, we find
\begin{equation}
\begin{aligned}
  k_a = M_a{}^b v_b   \, , \quad    k^a =   M^{ab} M^c{}_b k_c = \pi^{ab} k_b\,,  \quad  
   \varphi_a = 0 \,,  \quad \varphi^a = \llangle T^a, g^{-1} \dd g \rrangle \, .  
  \end{aligned} 
\end{equation}
Here $\varphi^a$ is dual to the $k_a$ which generate  right translations and $\pi :G \to \mathfrak{g}\wedge \mathfrak{g}$ defines   a Poisson bi-vector that obeys the analogue of a cocycle condition  making $G$ a Poisson-Lie group.

\section{Dressing cosets and generalised coset spaces}\label{sec:gencoset}
A dressing coset $M= \widetilde{G}\backslash\mathdsl{D} / H$ arises if a second isotropic subgroup $H$ is modded out from the coset $\widetilde{G}\backslash\mathdsl{D}$.   Let us now  reset notation  and define the generators of $\mathdsl{D}$ as $T_{\mathcal{A}}$.  We let the Lie algebra $\mathfrak{h}$ of $H$ be generated by  $T_\alpha$.  We let   $\mathfrak{d} = \mathfrak{h} + \mathfrak{k}$ and make a further splitting amongst the coset generators into $\mathfrak{k}= \mathfrak{p}+  \mathfrak{q}$ such that the pairing $\llangle \bullet   , \bullet \rrangle$ is  non-degenerate on $\mathfrak{q}$ and $\dim\mathfrak{q} = \dim \mathfrak{d} - 2 \dim \mathfrak{h} = 2 D$ is twice the dimension of the target space $M$.  We introduce now generators $T_A$ whose span gives  $\mathfrak{q}$ and $T^{\alpha}$ spanning  $\mathfrak{p}$, constructed so as to obey 
\begin{align}\label{eqn:pairingdoublecoset}
  \llangle T_\alpha , T_\beta \rrangle &= 0 \, ,  &
  \llangle T_\alpha , T^\beta \rrangle &= \delta_\alpha^\beta \, ,   &
  \llangle T_\alpha , T_B \rrangle &= 0 \, ,   &
  \llangle  T^\alpha , T_B \rrangle &= 0 \, .
\end{align}
We make an additional requirement: not only should $\mathdsl{D} / H$ be reductive such that $ \mathfrak{k}$ carries an action of $H$, but $\mathfrak{p}$ and $\mathfrak{q}$ themselves should further form two independent representations. This will be required so that our generalised frame fields have well defined properties under $H$ transformations and can be thought of as the generalised analogous of  reductive cosets. Hence, we call this property {\it generalised reductive}.

We now seek to construct a generalised frame field $E_A$ and an  $\mathfrak{h}$-valued  generalised  connection $\Omega^\alpha T_\alpha = \Omega^{\alpha B} E_B T_\alpha $   on $M= \widetilde{G}\backslash\mathdsl{D} / H$ obeying the frame algebra 
\begin{equation}\label{eqn:framealge2_sec3}
  \begin{aligned}
    \mathscr{L}_{E_A} E_B & = \mathcal{F}_{AB}{}^C E_C \equiv F_{AB}{}^C E_C + 2 \Omega^\alpha{}_{[A} F_{\alpha B]}{}^C E_C + F_{\alpha AB} \Omega^{\alpha} \, , \\
    \mathscr{L}_{\Omega^{\alpha}{}} E_B &=  (\Omega^{\beta\alpha} F_{\beta B }{}^C + F^\alpha{}_{B}{}^C) E_C + F_{\beta\gamma}{}^{\alpha} \Omega^{\beta}{}_B \Omega^\gamma\,,
  \end{aligned}
 \end{equation}
presented in the introduction. We additionally demand, in analogy to the pairing condition of eq.~\eqref{eqn:framealge_sec2}, that  
\begin{equation}\label{eqn:pairing}
  \langle E_A, E_B \rangle  = \llangle  T_A, T_B \rrangle \equiv \eta_{AB}  \, , \qquad  \langle \Omega^\alpha, E_B \rangle = \Omega^\alpha{}_B \,, \qquad  \langle \Omega^\alpha, \Omega^\beta \rangle = \llangle T^\alpha, T^\beta \rrangle + 2 \Omega^{(\alpha\beta)} \, , 
\end{equation}
such that $\Omega^\alpha{}_B= \Omega^{\alpha A} \eta_{AB}$. Additional constraints on the constituents of the algebra~\eqref{eqn:framealge2_sec3} arise from the Jacobi identity of the generalised Lie derivative. We defer this point to the next section, where we show there that $E_A$ and $\Omega^\alpha$ constructed in the following automatically satisfy these additional constraints.

In the following construction, we inherit the vectors fields defined in \eqref{eqn:ingredientsgenframe} (now renamed $k_{\mathcal{A}}$ like the generators), from which we form a restriction\footnote{To avoid burdensome notation with e.g. double over bars, we use in this section the over bar for further restriction to $M= \widetilde{G}\backslash\mathdsl{D} / H$, whereas in Section~\ref{sec:genpara} it denoted just restrictions to $\widetilde{G}\backslash\mathdsl{D}$.}  to the dressing coset
\begin{equation}
  \overline{k}_{\mathcal{A}} = k_{\mathcal{A}} - \iota_{k_{\mathcal{A}}} \mathscr{A}^\beta k_\beta \,.
\end{equation}
Here $\mathscr{A}$ denotes an $\mathfrak{h}$-valued flat connection that obeys $\iota_{k_\alpha}\mathscr{A}^\beta = \delta_\alpha^\beta$ and $L_{k_\alpha} \mathscr{A}^\beta = - F_{\alpha\gamma}{}^{\beta} \mathscr{A}^\gamma$. Suppose that we can choose (locally perhaps) the element $\mathdsl{g} = \widetilde{g} n h$ with $h \in H$ and $\widetilde{g} \in \widetilde{G}$ such that $n$ parametrises the double coset  $M= \widetilde{G}\backslash\mathdsl{D} / H$ then a canonical choice for this connection is $\mathscr{A}= h^{-1} d h$ which evidently obeys $\dd \mathscr{A} + \mathscr{A} \wedge \mathscr{A} =0 $. A subtlety comes when the gauge fixing $\mathdsl{g} = \widetilde{g} n h$  is not sufficient because some of the $H$ gauge transformations can be absorbed on a certain locus into compensating $\widetilde{G}$ transformations or, equivalently, when $H$ acts with fixed points on $\widetilde{G}\backslash\mathdsl{D}$. This will show up in singularities in the geometry as we illustrate with later examples. Furthermore, global properties may obstruct a globally flat connection. This case necessitates for the inclusion of localised vector multiplets associated as we discuss in Section~\ref{sec:sources} and demonstrate with examples.
 
The Lie bracket of the projected vector fields reads
\begin{equation}\label{eqn:kalgebradressing} 
  [\overline{k}_{\mathcal{A}}, \overline{k}_{\mathcal{B}}] = F_{\mathcal{A}\mathcal{B}}{}^{\mathcal{C}} \overline{k}_{\mathcal{C}} - 2 \iota_{k_{[\mathcal{A}}} \mathscr{A}^\delta F_{\delta | \mathcal{B}]}{}^{\mathcal{C}} \overline{k}_{\mathcal{C}}  \, . 
\end{equation}
Restricting the algebra~\eqref{eqn:kalgebradressing} to the index $A$ and the horizontal part of the Lie derivative, we obtain
\begin{equation}\label{eqn:liebrkilling}
  [ \overline{k}_A, \overline{k}_B ] = F_{AB}{}^C \overline{k}_C + 2 \Omega^{\delta}{}_{[A} F_{\delta B]}{}^C \overline{k}_C + F_{AB\gamma} \overline{k}^\gamma\,,
\end{equation} 
with
\begin{equation}\label{eqn:defOmega}
  \Omega^\alpha{}_B = - \iota_{k_B} \mathscr{A}^\alpha \qquad \text{and} \qquad \Omega^{\alpha\beta} = - \iota_{k^\beta} \mathscr{A}^\alpha\,,
\end{equation}
where the latter object has been defined for later purposes.
A comparison of the algebra eq.~\eqref{eqn:liebrkilling} and the desired frame algebra eq.~\eqref{eqn:framealge2_sec3} suggests that  
\begin{equation}\label{eqn:frameandconnectiondc}
  E_A = \overline{k}_A + \overline{\phi}_A \qquad \text{and} \qquad
  \Omega^\alpha = \overline{k}^\alpha + \overline{\phi}^\alpha\,.
\end{equation}
The differential forms $\phi_\mathcal{A}$ (with the restriction defined by $\overline{\phi}_{\mathcal{A}} = \phi_{\mathcal{A}} - \iota_{k_\beta} \phi_{\mathcal{A}} \mathscr{A}^\beta$) are $B$-shifted versions,
\begin{equation}
  \phi_{\mathcal{A}} = \varphi_{\mathcal{A}} - \iota_{k_{\mathcal{A}}} \mathscr{B}  \qquad \text{with} \qquad
  \mathscr{B} = \mathscr{A}^\beta\wedge\varphi_\beta  + \frac12 \iota_{k_\beta} \varphi_\gamma \mathscr{A}^\beta \wedge \mathscr{A}^\gamma \, , 
\end{equation}
of $\varphi_A$ which we introduced in eq.~\eqref{eqn:varphiA} of the last section. Because of this, $E_A$ and $\Omega^\alpha$ have a natural interpretation as coming directly from the reduction of the generalised frame field on $\widetilde{G} \backslash \mathdsl{D}$ after a particular gauge fixing. 
   The precise form of $\mathscr{B}$ ensures that   $\phi_\alpha = 0$ and that    $\iota_{k_\alpha} \phi_\mathcal{A} = \llangle T_\alpha, T_{\mathcal{A}} \rrangle$. This tells us that $\overline{\phi}_A = \phi_A$ and $\overline{\phi}^\alpha = \phi^\alpha - \mathscr{A}^\alpha$,  from which it is straightforward to see that the pairings of eq.~\eqref{eqn:pairing} hold. 

In order to show that the algebra of the reduced   vector fields in eq.~\eqref{eqn:liebrkilling} extends to the full generalised tangent space $T M \oplus T^ \star M$, we first note the two important properties of the $B$-field used to define $\phi_{\mathcal{A}}$ namely,    
\begin{equation}
  \iota_{k_\alpha} \mathscr{B} =   \varphi_\alpha \qquad \text{and} \qquad L_{k_\alpha}  \mathscr{B}  = 0\,. 
\end{equation}
Combining these with the result obtained previously that $ L_{k_{\mathcal{A}} 
}\varphi_{\mathcal{B}} = F_{ \mathcal{A}\mathcal{B}}{}^{\mathcal{C} } \varphi_{\mathcal{C}}$, one finds
\begin{equation}
    L_{\overline{k}_A} \overline{\phi}_B = F_{AB}{}^C \overline{\phi
   }_C +  F_{AB\gamma} \overline{\phi}^\gamma +    F_{AB\gamma} \mathscr{A}^\gamma+ \Omega^\gamma{}_A F_{\gamma B}{}^C \overline{\phi}_C - \iota_{k_B} L_{k_A} \mathscr{B} \, ,
\end{equation}
and
\begin{equation}
  \iota_{\overline{k}_A} ( H + \dd \mathscr{B} ) = \dd \phi_A +L_{k_A} \mathscr{B}\,.
\end{equation}
Both eventually give rise to
\begin{equation}\label{eq:formbit}
  L_{\overline{k}_A} \overline{\phi}_B - \iota_{\overline{k}_B} \dd \overline{\phi}_A - \iota_{\overline{k}_A} \iota_{\overline{k}_B}\mathscr{H} = F_{AB}{}^C \overline{\phi}_C + F_{AB \gamma} \overline{\phi}^\gamma + F_{AB \gamma} \mathscr{A}^\gamma  + 2 \Omega^\gamma{}_{[A} F_{\gamma B]}{}^C \overline{\phi}_C\, ,
\end{equation}
in which the closed three-form $\mathscr{H}= H + \dd \mathscr{B}$ has no vertical contributions, i.e. $\iota_{k_\alpha} \mathscr{H} = 0$. The LHS of this equation captures the one-form contribution of the $\mathscr{H}$-twisted generalised Lie derivative $\mathscr{L}_{E_A} E_B$ and up to the term $F_{AB\gamma} \mathscr{A}^\gamma$ the RHS has exactly the form dictated by the frame algebra~\eqref{eqn:framealge2_sec3}. To get rid of the additional term involving the connection, one remembers that both the vector fields $\overline{k}_A$ and $\overline{k}^\alpha$ as well as the corresponding one-forms, are still defined on $\widetilde{G}\backslash \mathdsl{D}$. Hence one has to push them forward/pull them back to the dressing coset. This is done by applying $\pi_*$ and $\sigma^*$, where $\pi$ projects onto the dressing coset, $\pi:\, \widetilde{G}\backslash\mathdsl{D} \rightarrow \widetilde{G}\backslash\mathdsl{D}/H$, and $\sigma$ is a section chosen such that $\sigma^* \mathscr{A} = 0$. Because $\mathscr{A}$ is a flat connection, a section with this property has to exist (at least patchwise). The procedure is essentially the familiar gauge fixing that removes extraneous degrees of freedom in a conventional coset construction. Concluding, eq.~\eqref{eq:formbit} with the vector part eq.~\eqref{eqn:liebrkilling} ensures that the first part of the frame algebra eq.~\eqref{eqn:framealge2_sec3} holds for the generalised frame field
\begin{equation}
  E_A = \pi_* \overline{k}_A + \sigma^* \overline{\phi}_A \qquad
    \text{and} \qquad
  \Omega^\alpha = \pi_* \overline{k}^\alpha + \sigma^* \overline{\phi}^\alpha\,.
\end{equation}

In order to obtain the second generalised Lie derivative $\mathscr{L}_{\Omega^\alpha} E_B$ in eq.~\eqref{eqn:framealge2_sec3}, a short calculation verifies that
\begin{equation}\label{eqn:dOmegaantisym}
  2 \iota_{\overline{k}_[A} \dd \Omega^{\gamma}{}_{B]} = - F_{\delta\epsilon}{}^{\gamma} \Omega^{\delta}{}_A \Omega^{\epsilon}{}_B -  F_{AB}{}^{\gamma} + F_{AB}{}^D \Omega^{\gamma}{}_D + F_{AB\delta} \Omega^{\gamma\delta} + 2 F_{[A|\delta}{}^E \Omega^{\delta}{}_{B]} \Omega^{\gamma}{}_E
\end{equation} 
and then we note that
\begin{equation}
  \mathscr{L}_{\Omega^\alpha} E_B = \left( \Omega^\alpha{}_D \mathcal{F}^D{}_{BC} -2\iota_{\overline{k}_{[B}} \dd \Omega^\alpha{}_{C]} \right) E^C 
\end{equation} 
holds.  After some careful rearrangement, exploiting the anti-symmetry of $F_{\mathcal{A}\mathcal{B}\mathcal{C}}$, one finds many terms on the right hand side of the above combine or cancel to recover the stated   frame algebra in \eqref{eqn:framealge2_sec3}.

\subsection{Generalised torsion, curvature and intrinsic curvature}
We would now like to evaluate various properties of generalised coset spaces constructed above.   At this juncture we point out that there are two common approaches to describing the frame fields.  One can, as we have thus far, construct frame fields that furnish an algebra under the $H$-twisted generalised Lie derivative. This avoids working patchwise picking a local representative for a potential for $H$.  An alternative, however, is to exactly do that, i.e. work locally with the untwisted generalised Lie derivative and absorb the corresponding (local) potential $H= \dd B$  into the frame field. As a scalar, $\Omega^\alpha{}_B$ is not affected by this transformation. Many of the formulae regarding curvatures are more readily found in the literature with this later convention. Since  this  also  lends itself to simpler expressions in the explicit examples we shall later consider,  we adopt it henceforth (using the same symbol $E_A$ for the frames to avoid clutter).

Our first step is   to introduce a covariant derivative 
\begin{equation}
  \nabla_I E_A{}^J = \partial_I E_A{}^J - \Omega_{IA}{}^B E_B{}^J + \Gamma_{IK}{}^J \,,E_A{}^K
\end{equation}
in which  $\partial_I = \begin{pmatrix} 0 \, ,& \partial_i \end{pmatrix}$ is  a partial derivative. Here we use the abbreviation
\begin{equation}
  \Omega_{AB}{}^C = \Omega^{\delta}{}_A  F_{\delta B}{}^C
\end{equation}
and switch between flat indices, $A,B,C,\dots$, and curved indices, $I,J,K,\dots$, with the generalised frame $E_{A}{}^I$ and its inverse  transpose  $E^B{}_I$.  The first feature of this connection is that  $\nabla_{I} \eta_{AB} = 0 $ since $\eta_{AB}$ is both constant and  invariant under the $H$-action generated by the spin-connection. 

Imposing the vielbein postulate, $\nabla_I E_A{}^J = 0$, allows one to express the generalised Christoffel symbols
\begin{equation}
  \Gamma_{IJK} = \partial_I E^A{}_J E_{AK} + \Omega_{IJK}\,,
\end{equation}
in terms of the generalised frame field $E$ and the spin connection $\Omega$. In particular, we can make use of this fact  to compute the generalised torsion \cite{Hohm_2013}
\begin{equation}
  T_{ABC} = 3 \Gamma_{[ABC]} = - 3 E_{[A}{}^I \partial_I E_{B}{}^J E_{C]J} + 3 \Omega_{[ABC]}\,.
\end{equation}
On the other hand, the first generalised Lie derivative in \eqref{eqn:framealge2_intro} of a generalised coset space results in
\begin{equation}\label{eqn:gentorsionconstr}
  \mathscr{L}_{E_A} E_B^I E^C{}_I = 
  3 E_{[A}{}^I \partial_I E_B{}^J E_{C]J} = F_{ABC} + 3 \Omega_{[ABC]}\,.
\end{equation}
As a result of eq.~\eqref{eqn:gentorsionconstr}, the generalised torsion in flat indices is constant, $T_{ABC} = - F_{ABC}$ and just governed by some of the structure coefficients of the Lie algebra $\mathfrak{d}$. This situation is in perfect analogy with the torsion of a reductive coset space. When $F_{ABC}\equiv 0$, as will be the case in the examples considered later, we have a natural notion of a {\em generalised symmetric space}, a property one expects to play a key role for integrability of the string world sheet theory. 

In the same vein, one calculates the generalised curvature \cite{Hohm_2013}
\begin{equation}
  R_{IJKL} = 2 \partial_{[I} \Gamma_{J]KL} + 2 \Gamma_{[I|ML} \Gamma_{|J]K}{}^M + \frac12 \Gamma_{MIJ} \Gamma^M{}_{KL} + (IJ) \leftrightarrow (KL)\,,
\end{equation}
whose flat version evaluates to
\begin{equation}
  \begin{aligned}
    R_{ABCD} &= 2 E_{[A}^I \partial_I \Omega_{B]CD} - 2 \Omega_{[A|DE} \Omega_{B]C}{}^E - ( F_{ABE} + 2 \Omega_{[AB]E} ) \Omega^E{}_{CD} - \\
    &\frac12 \Omega_{EAB} \Omega^E{}_{CD} + (AB) \leftrightarrow (CD)\,.
  \end{aligned}
\end{equation}
At this point it is useful to note that the frame algebra of the generalised coset implies the relation \eqref{eqn:dOmegaantisym} which can be equally written as
\begin{equation}\label{eqn:DOmega}
  E_{[A|}{}^I \partial_I \Omega_{|B]CD} = \Omega_{[A|C}{}^E \Omega_{|B]DE} + \left( \frac12 F_{ABE} + \Omega_{[AB]E} \right) \Omega^{E}{}_{CD} - \frac12 F_{AB}{}^{\epsilon} F_{\epsilon CD}\, ,
\end{equation}
which guarantees that the generalised curvature simplifies to 
\begin{equation}
  R_{ABCD} = - F_{AB}{}^\epsilon F_{CD\epsilon} - F_{AB\epsilon} F_{CD}{}^\epsilon - F_{AB\delta} F_{CD\epsilon} \llangle T^\delta, T^\epsilon \rrangle \,.
\end{equation}
Hence we conclude that the generalised torsion and curvature of a generalised coset are completely fixed by the structure coefficients of the underlying Lie algebra $\mathfrak{d}$. This result might seem surprising, for one crucial difference between geometry and generalised geometry is that for a torsion free connection the generalised Riemann tensor can not be completely fixed using the metric, $B$-field and dilaton. There remain undetermined components which however do not affect the generalised Ricci tensor and scalar. The reason for this feature is that the metric, $B$-field and dilaton only fix an $O(d,d)$ frame up to a local double Lorentz transformation valued in $O(d)\times O(d)$. The construction we present, however, also singles out a particular double Lorentz frame and thus determines the connection and curvature completely. This point is particularly important if $E_A$ acts on spinors which transform non-trivially under the double Lorentz group. Hence the constructed generalised frame fields not only contains information about the NS/NS sector of type II string theory but also about the R/R sector. In particular this allows to extract the R/R sector transformation rules for the dressing coset construction along the lines of \cite{Hassler:2017yza}.

There are two simple checks of the presented results. First, the closure of the frame algebra in \eqref{eqn:framealge2_intro} requires that
\begin{equation}
  E_{[A}^I \partial_I \mathcal{F}_{BCD]} = \frac{3}{4} \mathcal{F}_{[AB}{}^{E} \mathcal{F}_{CD]E}\,,
\end{equation}
which is indeed satisfied due to \eqref{eqn:DOmega}. Second, we can check the Bianchi identity \cite{Hohm_2013}
\begin{equation}\label{eqn:genbianchi}
  3 R_{[ABCD]} = 4 \nabla_{[A} T_{BCD]} + 3 T_{[AB}{}^E T_{CD]E} \, , 
\end{equation}
for the torsionful generalised connection $\nabla$. In order to prove it, we first note that 
\begin{equation}
  \nabla_A T_{BCD} = 0\,,
\end{equation}
because $\nabla_A$ acts on constant tensors just by the $H$-adjoint action but $T_{ABC} = - \llangle [ T_A, T_B ] , T_C \rrangle$ is ad-invariant. Thus, the Bianchi identity \eqref{eqn:genbianchi} reduces to the Jacobi identity on the Lie algebra $\mathfrak{d}$. In the same vein, one derives
\begin{equation}
  \nabla_A R_{BCDE} = 0\,,
\end{equation}
after taking into account that $\eta_{AB}$ is covariantly constant. 

Typically we are only interested in the action of covariant derivatives on objects that are invariant under the adjoint action of $H$. It is then natural to subtract from the torsion the portion that depends on the adjoint action of $H$ by defining the intrinsic torsion
\begin{equation}
  T^{\mathrm{int}}_{ABC} = T_{ABC} - T_{ADE} P_{\mathfrak{h} }^{DE}{}_{BC}\,,
\end{equation}
where $P_{ \mathfrak{h} }$ projects onto the adjoint representation of the $\mathfrak{h}$. Because the projector is covariantly constant with respect to $\nabla$, the intrinsic torsion is a singlet under $H$ action (or equally it is covariantly constant too). In general, tensors in flat indices that are constant and $H$-invariant are covariantly constant on $M$. This is a considerable advantage of generalised cosets because invariant tensors are completely fixed by the Lie algebra $\mathfrak{h}$ and the  problem of solving complicated PDEs is circumvented. Any parallel transport around a closed loop in $M$ is infinitesimally mediated by the covariant derivative $\nabla$. Hence monodromies for loops in one patch always have to be valued in $H$. Transitioning between patches also only involved $H$-transformations which originate from patching the flat connection $\mathscr{A}$. Thus, we conclude that the generalised structure group $G_{\mathrm{S}}$ of $M$ is $H$. Now the list of required properties for a consistent truncation given in the introduction is complete.

The same idea also applies to scalar densities like $e^{-2 d}$ which governs the generalised dilaton $d = \Phi - 1/4 \log \det g$, where $g$ denotes the metric. This quantity is only covariantly constant if
\begin{equation}
  \Gamma_{JI}{}^J = - 2 \partial_I d
\end{equation}
holds. This relation fixes the exterior derivative of the dilaton $\Phi$, once the metric is known via
\begin{equation}
  \dd \Phi = \frac14 \dd (\log \det g) - \frac12 \Gamma_{Ji}{}^J \dd x^i\,.
\end{equation}
Alternatively, one can encode the dilaton in the fluxes
\begin{equation}
  \mathcal{F}_A = E^{B I} \partial_I E_B{}^J E_{AJ} + 2 E_A{}^I \partial_I d = \Omega^B{}_{BA} \,.
\end{equation}
Together with $\mathcal{F}_{ABC}$ they form the natural objects in the flux formulation of double field theory \cite{Geissbuhler:2013uka}. In terms of them the supergravity field equations and action for the bosonic sector have a simple form.

A further simplification arises for generalised symmetric spaces with an $H$-invariant generalised metric $\mathcal{H}_{AB}$. In this case $\nabla_A$ is a generalised Levi-Civita connection and the supergravity field equations for the NS/NS sector can be written in terms of the generalised Ricci tensor and the generalised Ricci curvature,
\begin{equation}
  R_{AB} = \overline{P}^{CD} R_{ACBD}
    \qquad \text{and} \qquad
  R = \overline{P}^{AB} R_{AB} = - R_A{}^A\,, 
\end{equation}
as
\begin{equation}
  \widehat{R}_{AB} = \overline{P}_{(A}{}^C P_{B)}{}^D R_{CD} = 0
    \qquad \text{and} \qquad
  R = 0
\end{equation}
with the projectors $\overline{P}_{AB} = 1/2 ( \eta_{AB} - \mathcal{H}_{AB} )$ and $P_{AB} = 1/2 ( \eta_{AB} + \mathcal{H}_{AB} )$. In contrast to general relativity not all components of the generalised Ricci tensor $R_{AB}$ contribute to the field equations but only a particular projection, here denotes as $\widehat{R}_{AB}$. Using that the generalised metric is $H$-invariant, we obtain
\begin{equation}
  \widehat{R}_{AB} = \overline{P}_{(A}{}^C P_{B)}{}^D \kappa_{CD}
\end{equation}
where $\kappa_{AB} = - F_{A\mathcal{C}}{}^{\mathcal{D}} F_{B\mathcal{D}}{}^{\mathcal{C}}$ denotes the canonical Killing metric restricted to the generators $T_A$. If this metric is non-degenerate and has appropriate signature, one can choose $\mathcal{H}_{AB} = \kappa_{AB}$ to obtain $\widehat{R}_{AB} = 0$ and $R = - \dim M$. This is the natural generalisation of a maximally symmetric space.

\subsection{Localised sources and extended generalised tangent space   }\label{sec:sources}
A crucial part of the construction we presented above is played by the flat connection $\mathscr{A}$. In general, however, there are cases where it is not possible to define a connection which is flat everywhere on $M$. A simple example, which we will explore in full detail in the next section, is an $H=U(1)_\mathrm{A}$ connection which is patched by a gauge transformation on a contractible cycle. Stokes' theorem implies that there has to be a non-vanishing field strength $\mathscr{F} = \dd \mathscr{A}$ when this cycle collapses to a point $P$. Assume further that the vector field $k^\alpha$ (in this case since $\dim H=1$, we have that a single vector field, i.e. $\alpha=1$) generates this cycle. Then the distinguished point $P$ with non-vanishing field strength $\mathscr{F}$ is exactly the fixed point of the corresponding $U(1)_\mathrm{V}$ acting on $\widetilde{G}\backslash \mathdsl{D}$. In this simple example both $U(1)_\mathrm{A}$ and $U(1)_\mathrm{V}$ are isotropic subgroups and thus we can alternatively study a generalised coset $\widetilde{M}$ with $H=U(1)_\mathrm{V}$. Now the fixed point $P$ of the $H$-action results in a singularity on $\widetilde{M}$. One can understand this exchange of $U(1)_\mathrm{A}$ and $U(1)_\mathrm{V}$ as a T-duality transformation. Hence we conclude that one situation in which $\mathscr{F}$ vanishes everywhere except for a distinguished point is related to the resolution of a singularity by T-duality. Of course the process for a non-abelian $H$ is more involved. However, we take this simple example as a motivation to discuss how dropping the constraint of a flat connection affects the construction of generalised cosets.

Most notably, the Lie bracket in eq.~\eqref{eqn:liebrkilling} receives an additional contribution
\begin{equation}
  [\overline{k}_A, \overline{k}_B ] = F_{AB}{}^C \overline{k}_C  + 2 \Omega^\delta{}_{[A} F_{\delta B]}{}^C \overline{k}_C + F_{AB\gamma} \overline{k}^\gamma + \iota_{\overline{k}_A} \iota_{\overline{k}_B} \mathscr{F}^\gamma k_\gamma
\end{equation}
from the field strength of $\mathscr{A}$ which is encoded in
\begin{equation}
  \mathscr{F}_{AB}{}^\gamma = \iota_{\overline{k}_A} \iota_{\overline{k}_B} \mathscr{F}^\gamma
    \qquad \text{with} \qquad 
  \mathscr{F}  = \mathscr{F}^\alpha T_\alpha = d\mathscr{A} + \mathscr{A}\wedge \mathscr{A}\, .
\end{equation}
The analogous contribution in eq.~\eqref{eq:formbit} for the one-form part of the generalised Lie derivative remains unchanged. Thus, there is no obvious counterpart to the additional term for the vectors which could give rise to an object on the generalised tangent space. This problem is fixed by decomposing the $H$-flux into
\begin{equation}\label{eqn:Hdecompcalphabeta}
  H = H_0 + {\omega}_{CS} \ , \quad   {\omega}_{CS} = c_{\alpha\beta} \left( d\mathscr{A}^\alpha \wedge \mathscr{A}^\beta + \frac{1}{3} F_{\gamma \delta}{}^\alpha   \mathscr{A}^\beta \wedge \mathscr{A}^\gamma \wedge \mathscr{A}^\delta  \right)   \,,
\end{equation}
where $c_{\alpha\beta}$ is a symmetric, non-degenerate, ad-invariant pairing on $\mathfrak{h}$. Note that,  unlike $\mathscr{F}$, the Chern-Simons form is not horizontal, instead it obeys
\begin{equation}
  \iota_{k_\alpha} {\omega}_{CS} = c_{\alpha\beta} \dd \mathscr{A}^\beta \, , \quad \iota_{k_\alpha} \iota_{k_\beta} {\omega}_{CS} = F_{\alpha \beta}{}^\delta  c_{\delta\gamma}  \mathscr{A}^\gamma\, , \quad \iota_{\bar{k}_A} \iota_{\bar{k}_B} {\omega}_{CS} =  \mathscr{F}_{AB}{}^\gamma c_{\gamma\delta}  \mathscr{A}^\delta\,,
\end{equation}
and crucially 
\begin{equation}
  d{\omega}_{CS} = c_{\alpha\beta}  \mathscr{F}^\alpha\wedge  \mathscr{F}^\beta \,,
\end{equation}
ensures that it is invariant i.e. $L_{k_\alpha}  {\omega}_{CS} = 0$. We will now consider twisting the Courant bracket with the H-flux $\mathscr{H}_0 = H_0 + \dd \mathscr{B}$ which differs from the $\mathscr{H}  = H  + \dd \mathscr{B}$ used previously by the above Chern-Simons three-form contribution.
 
This allows us to substitute eq.~\eqref{eq:formbit} with
\begin{equation}\label{eq:formbithet}
  L_{\overline{k}_A} \overline{\phi}_B - \iota_{\overline{k}_B} \dd \overline{\phi}_A - \iota_{\overline{k}_A} \iota_{\overline{k}_B}\mathscr{H}_0 = F_{AB}{}^C \overline{\phi}_C + F_{AB \gamma} \phi^\gamma + 2 \Omega^\gamma{}_{[A} F_{\gamma B]}{}^C \overline{\phi}_C + \mathscr{F}_{AB}{}^\gamma c_{\gamma\delta} \mathscr{A}^\delta \, ,
\end{equation}
such that RHS contains the appropriate one-form counterpart to $\mathscr{F}_{AB}{}^\gamma k_\gamma$.  All together we now consider the components  
 \begin{equation}
   E_A = M_A{}^B ( \overline{k}_{B} + \overline{\phi}_B ) \ , \quad \Omega^\alpha = M^\alpha{}_\beta ( \overline{k}^{\beta} + \phi^\beta ) \, ,   \quad    E_\alpha = M_\alpha{}^\beta ( k_\beta + c_{\beta\gamma} \mathscr{A}^\gamma ) \, , 
\end{equation}
of the generalised frame field with the additional pairings
\begin{equation}
  \langle E_\alpha, E_\beta \rangle = 2 c_{\alpha\beta}\,, \quad 
  \langle E_\alpha, E_B \rangle = 0  \qquad \text{and} \qquad
  \langle E_\alpha, \Omega^\beta \rangle = \delta_\alpha{}^\beta\,.
\end{equation}
$M_{\mathcal{A}}{}^{\mathcal{B}}$ denotes the adjoint $H$ action on the Lie algebra $\mathfrak{h}$ which is defined by $M_{\mathcal{A}}{}^{\mathcal{B}} t_{\mathcal{B}} = h^{-1} t_{\mathcal{A}} h$ where $h$ arises from the decomposition $\mathdsl{d} = \widetilde{g} n h$ with $n\in \widetilde{G}\backslash\mathdsl{D}/H$. Taking into account the new components of the generalised frame field the original frame algebra is extended to
\begin{equation}
  \begin{aligned}
    \mathscr{L}_{E_A} E_B &= \mathcal{F}_{AB}{}^C E_C + \mathscr{F}_{AB}{}^\gamma E_\gamma \,, \\
    \mathscr{L}_{E_\alpha} E_\beta & = - F_{\alpha\beta}{}^\gamma E_\gamma \,, \\
    \mathscr{L}_{E_\alpha} E_B &= 2 \mathscr{F}_B{}^{C \delta} c_{\delta\alpha} E_C = - \mathscr{L}_{E_B} E_\alpha\,,
  \end{aligned}
\end{equation}
with an $\mathscr{H}_0$ twisted generalised Lie derivative. These three equations can be elegantly combined by considering the generalised Lie derivative \cite{Hohm:2011ex,Coimbra:2014qaa}
\begin{equation}
  \begin{aligned}
    \widehat{\mathscr{L}}_{\widehat{U}} \widehat{V} = &[u,v] + L_u \nu - \iota_v d \mu - \iota_u \iota_v \mathscr{H}_0 + 2 c( t, \iota_u \mathscr{F} ) - 2 c( s, \iota_v \mathscr{F} ) + 2 c( D s, t )   \\
    & \qquad + D_u t - D_v s - [s, t] + \iota_u \iota_v \mathscr{F}\,,
  \end{aligned}
\end{equation}
on the extended generalised tangent bundle $T M \oplus T^\star M \oplus \mathrm{ad}\, \mathfrak{h}_0$ with generalised vectors $\widehat{U} = u + \mu + s$ and $\widehat{V} = v + \nu + t$ where $s, t \in \mathfrak{h}$, the pairing $c(t_\alpha, t_\beta) = c_{\alpha\beta}$ and the (gauge) covariant derivate $D = \dd + \mathscr{A} \wedge$. The corresponding generalised frame field has the components
\begin{equation}
  \widehat{E}_A = E_A \qquad \text{and} \qquad \widehat{E}_\alpha = t_\alpha + c_{\alpha\beta} A^\beta\,.
\end{equation}

A particular feature of this construction is that the $H$-flux $\mathscr{H}_0$ is not closed anymore but instead is governed by
\begin{equation}
  \dd \mathscr{H}_0 = - c( \mathscr{F} \overset{\wedge}{,} \mathscr{F} )
\end{equation}
which indicates magnetic sources for $H$. As mentioned earlier, our construction is related to a gauging procedure in two-dimensional $\sigma$-models. In this context the pairing $c_{\alpha\beta}$ which was introduced in the decomposition \eqref{eqn:Hdecompcalphabeta} has a natural interpretation. Assume that we are in the region of $M$ where $\mathscr{F}$ vanishes and $\mathscr{H}_0$ is closed. Furthermore, it can be easily verified that
\begin{equation}
  L_{k_\alpha} \mathscr{H}_0  = 0 \qquad \text{and} \qquad
  \iota_{k_\alpha} \mathscr{H}_0 = - c_{\alpha\beta} \dd A^\beta
\end{equation}
holds  everywhere on $M$. On the world sheet, the generalised coset is implemented by a gauged $\sigma$-model whose classical gauge anomaly is given by
\begin{equation}
  \delta_{\lambda} S = \int_\Sigma c_{\alpha\beta} \lambda^\alpha \mathscr{A}^\beta \,.
\end{equation}
One might worry that this gauge anomaly might indicate a problem, but in fact it only emphasise the point that the additional degrees of  freedom which originate for the vectors of the extended generalised tangent space are required to cancel the anomaly.

\section{Examples}\label{sec:ns5example} 
As an explicit example for the techniques introduced in the last two sections, we first discuss the generalised parallelisable space $G_{\mathrm{diag}} \backslash (G \times G)$. The group elements of $\mathdsl{D}$ are written as $\mathdsl{g} = (g_{\mathrm{L}}, g_{\mathrm{R}})$, where $g_{\mathrm{L}}$ and $g_{\mathrm{R}}$ are two elements of the Lie group $G$. Elements of $\widetilde G=G_{\mathrm{diag}} $ are parameterised by $(\tilde g, \tilde g)$, $\tilde g \in G$ and as coset representatives we choose $m = (g, e)$ (here $e$ denotes the unit element of $G$). Assume that the Lie group $G$, generated infinitesimally by $t_a$, is endowed with an ad-invariant, non-degenerate pairing
\begin{equation}
  \prec t_a, t_b \succ = \kappa_{ab}\,,
\end{equation}
then one has the defining relations for the vectors $k_{\mathrm{L}/\mathrm{R}}$ generating respectively left and right actions 
\begin{equation}
  \iota_{k_{\mathrm{L} a}} \dd g g^{-1} = t_a
    \qquad \text{and} \qquad
  \iota_{k_{\mathrm{R} a}} g^{-1} \dd g = t_a\,.
\end{equation}
Defined like this, the Killing vectors indeed give rise to the Lie algebra $\mathfrak{g} \oplus \mathfrak{g}$:
\begin{equation}
  [ k_{\mathrm{L}_a}, k_{\mathrm{L}_b} ] = -f_{ab}{}^c k_{\mathrm{L}_c} \,, \qquad
  [ k_{\mathrm{R}_a}, k_{\mathrm{R}_b} ] =  f_{ab}{}^c k_{\mathrm{R}_c} \qquad \text{and} \qquad
  [ k_{\mathrm{L}_a}, k_{\mathrm{R}_b} ] = 0\,,
\end{equation}
where $f_{ab}{}^c$ denote the structure coefficients of $\mathfrak{g}$. With the pairing
\begin{equation}
  \llangle ( t_a, t_b ), ( t_c, t_d ) \rrangle = \prec t_a, t_c \succ - \prec t_b, t_d \succ \,,
\end{equation}
the one-forms
\begin{equation}
  \varphi_{\mathrm{L} a} = \frac12 \prec t_a, g^{-1} \dd g \succ\,,
    \qquad
  \varphi_{\mathrm{R} a} = - \frac12 \prec t_a, \dd g g^{-1} \succ\,,
\end{equation}
and the $H$-flux
\begin{equation}\label{eqn:HfluxGxG}
  H = -\frac1{12} \prec g^{-1} \dd g, [ g^{-1} \dd g, g^{-1} \dd g ] \succ \,.
\end{equation}

In the following, we use these results to study dressing cosets based on $H=U(1)$ and $G=SU(2)$ which give a gauged WZW realisation \cite{Bardacki:1990wj} of the parafermionic CFT \cite{Fateev:1985mm}.    After this we choose $G=SU(2) \times SL(2,\mathbb{R})$ quotiented by a $U(1)\times U(1)$ to describe the transversal space formed by NS5-branes on a circle \cite{Sfetsos:1998xd}. For this purpose it is useful to note that $H$ with elements $(h_\mathrm{L}, h_\mathrm{R})$ acts as $g \rightarrow h_\mathrm{L}^{-1} g h_\mathrm{R}$ on the coset $G\backslash (G\times G)$. 
 
\subsection{Parafermions and their deformations}
Let us first consider $G=SU(2)$ whose elements can be embedded into $\mathbb{R}^4$ using the coordinates 
\begin{equation}
  \begin{aligned}
  x_1 &= \cos \theta \cos \widetilde\phi &
  x_2 &= \cos \theta \sin \widetilde\phi &
  x_3 &= \sin \theta \sin \phi &
  x_4 &= \sin \theta \cos \phi\,.
  \end{aligned}
\end{equation}
There are two isotropic $U(1)$ which one can gauge to obtain the dressing coset. They are parameterised by $\phi$ and $\widetilde\phi$ respectively. To see this explicitly, it is convenient to write the corresponding group elements as
\begin{equation}
  g = e^{i \sigma_3 ( \widetilde\phi - \phi )/2} e^{i \sigma_1 \theta} e^{i \sigma_3 ( \widetilde\phi + \phi )/2}\,,
\end{equation}
where $\sigma_\alpha$ denotes the Pauli matrices. The two resulting $U(1)$ are then given by 
\begin{equation}
  h_{\widetilde{\phi}} = ( e^{- i \sigma_3 \widetilde\phi/2}, e^{i \sigma_3 \widetilde\phi/2} ) \quad \text{(axial)}
    \qquad \text{and} \qquad
  h_\phi = ( e^{i \sigma_3 \phi/2}, e^{i \sigma_3 \phi/2} ) \quad \text{(vectorial)}\,.
\end{equation}
If we further fix the generators $t_\alpha = i/2 \sigma_\alpha$,  and the pairing $\prec t_\alpha, t_\beta \succ = - k \, \mathrm{Tr}(t_a t_b)$, eq.~\eqref{eqn:HfluxGxG} gives rise to
\begin{equation}
  H = k \sin 2\theta \,\dd \theta \wedge \dd \phi \wedge \dd \widetilde\phi\,.
\end{equation}

\subsubsection*{Axial gauge}
Next, we gauge the axial $U(1)_{\widetilde{\phi}}$. The canonical choice for a connection would be just $\mathscr{A} = h_{\widetilde{\phi}} \dd h^{-1}_{\widetilde{\phi}}$, however all other flat connections are admissible, too. It will turn out that $\mathscr{A}$ controls the monodromy of the generalised frame. For the setup we study here, this monodromy is controlled by two integers, $l$ and $\widetilde{l}$ with $l \widetilde{l} = k$. Hence, we will use the connection 
\begin{equation}\label{eqn:Aparaferm}
  \mathscr{A} = t_{\widetilde{\phi}} (l \dd \phi + \widetilde{l} \dd \widetilde{\phi}) \,,
\end{equation}
to construct the generalised coset. Here $t_{\widetilde{\phi}}$ denotes the generator of $U(1)_{\widetilde{\phi}}$ which corresponds to the vector field $k_{\widetilde{\phi}}=\widetilde{l}^{-1} \, \partial_{\widetilde{\phi}}$. Furthermore, there is the dual vector field, $k_\phi = l^{-1} \,\partial_\phi$ and the corresponding generator $t_\phi$. All of them are completely fixed by the connection~\eqref{eqn:Aparaferm}: the normalisation of $k_{\widetilde{\phi}}$ is determined by $\iota_{k_{\widetilde{\phi}}} \mathscr{A}=1$. Eventually, the duality condition $\llangle t_{\widetilde{\phi}}, t_\phi \rrangle = 1$ fixes the rest.

Applying the procedure outline in the last section, with this choice we obtain the generalised frame field
\begin{equation}
\mathscr{E}_L \equiv E_{\mathrm{L}1} + i  E_{\mathrm{L}2}= \frac12 e^{i \phi_+} \left (\partial_\theta + i \cot (\theta) \partial_\phi - k \dd \theta -i k \tan (\theta) \dd \phi \right) \,,
\end{equation}
while the right combination can be obtained replacing $\phi_+ \rightarrow \phi_-$ and $k \rightarrow - k$, with $\phi_\pm = (1 \pm l/\widetilde{l}) \phi$ and the generalised connection $\Omega = t_{\widetilde{\phi}} \, ( k_\phi - l \dd \phi )$. It is straightforward to check that they satisfy the frame algebra~\eqref{eqn:framealge2_intro}. Using the $H$-invariant generalised metric $\mathcal{H}_{\mathrm{LL}ab} = \mathcal{H}_{\mathrm{RR}ab} = \prec t_a, t_b \succ$, one obtains the metric and dilaton
\begin{equation}\label{eqn:metricdilatonparaferm} 
  \dd s^2 = k ( \dd \theta^2 + \tan^2 \theta \,\dd \phi^2 )\,, \qquad
  e^\Phi = g_{\mathrm{s}} \frac1{\cos\theta} \,.
\end{equation}
Although the assignments $l$ and $\tilde{l}$ drop out of the metric they are retained in the frame fields and play an important role in determining periodicities of coordinates as we shall see shortly. For $\theta = \pi/2$ the $H$ action has a fixed point and the theory is strongly coupled.

There are more choices for an $H$-invariant generalised metric, which can be constructed as follows: first, we determine the commutant subgroup of $O(2,2)$ and $H$ which is $SO(3)$. An $SO(2)$ subgroup of this $SO(3)$ is also a subgroup of $O(2)\times O(2)$ and hence it will not contribute to the generalised metric. We conclude that covariantly constant generalised metrics are parameterised by the coset $SO(3)/SO(2)$ with the coordinates $\alpha$ and $\beta$. More specifically, the generalised 
\begin{equation}\label{eq:Hforlambda}
  \begin{aligned}
  \mathcal{H}_{\mathrm{L}_a \mathrm{L}_a} &= \mathcal{H}_{\mathrm{R}_a \mathrm{R}_a} = \frac{k}{4} \frac{2 + 4 \alpha^2 (1 + \alpha^2) + \beta^2}{ 1 + 2 \alpha^2} \ , \\ \mathcal{H}_{\mathrm{L}_1 \mathrm{R}_1}  &= - \mathcal{H}_{\mathrm{L}_2 \mathrm{R}_2}  = \frac{k}{4} \frac{ 4 \alpha^2 (1 + \alpha^2) + \beta^2}{ 1 + \alpha^2 } \,, \\
   \mathcal{H}_{\mathrm{L}_1 \mathrm{R}_2} &= \mathcal{H}_{\mathrm{L}_2 \mathrm{R}_1} = \frac{k \beta}{2 + 4 \alpha^2}\,,
  \end{aligned}
\end{equation}
the resulting target space belongs to an integrable $\sigma$-model with vanishing  $B$-field which is discussed in \cite{SfetsosSiamposThompson}. Choosing $\beta=0$ and $\alpha = \sqrt{\lambda/(1-\lambda)}$ the metric becomes 
\begin{equation}
\dd s^2_{\lambda}=  k \frac{1- \lambda}{1+\lambda} ( \dd \theta^2 + \tan^2 \theta \,\dd \phi^2 )  +  k \frac{ 4 \lambda}{1-\lambda^2} ( \cos\phi \, \dd \theta - \sin \phi \tan\theta\, \dd  \phi )^2 \, .
\end{equation}
This metric is recognised as the axial-$\lambda$-deformation \cite{Sfetsos:2013wia,Driezen:2019ykp} of eq.~\eqref{eqn:metricdilatonparaferm}\footnote{This can be obtained alternatively using either the asymmetric gauging techniques of \cite{Driezen:2019ykp} or by considering the (conventional) vector gauged $\lambda$-model  and doing the coordinate change $\widetilde\phi \to \frac{\pi}{2}- \phi$ (corresponding to gauging with an inner-automorphism in \cite{Driezen:2019ykp}).}.

Let us now assume that we restrict the domain of $\phi$ to $0\le \phi < 2\pi / l$ which results for $l>1$ in a $\mathbb{Z}_l$ orbifold singularity $\theta=0$. In this case the generalised frame field exhibits the monodromy
\begin{equation}\label{eqn:monodromy}
  \mathscr{M}_A{}^B = \langle E_A(\phi=0) , E^B(\phi=2\pi/l) \rangle =
    \left(e^{2 \pi ( t_{\widetilde{\phi}} - t_\phi )}\right)_A{}^B\,,
\end{equation}
which is valued in both $U(1)_{\widetilde{\phi}}$ and $U(1)_\phi$. In fact the connection \eqref{eqn:Aparaferm} is fixed by requiring this particular monodromy. Its $U(1)_{\widetilde{\phi}}$ contributions arises directly from the connection by considering the integral
\begin{equation}
  e^{\oint_{S^1_\phi} \mathscr{A}} = e^{2 \pi t_{\widetilde{\phi}}}\,,
\end{equation}
where $S^1_\phi$ is a circle tracing out $\phi$ at a fixed value of $\theta$. This circle is the boundary of a disc $D^2$ centered around $\theta=0$. Using Stokes' theorem, one can alternatively write
\begin{equation}
  e^{\oint_{D^2} \mathscr{F}} = e^{2 \pi t_{\widetilde{\phi}}}.
\end{equation}
The radius of the disc does not affect the result, so one is able to contract it to the origin at $\theta=0$. This implies that there is $\mathscr{F}$-flux localised at the origin and we have to extend the generalised tangent bundle there with an additional vector along the lines of Section~\ref{sec:sources}.

\subsubsection*{Vectorial gauge}
More light is shed on the strongly coupled region with $\theta=\pi/2$ by considering the T-dual configuration which arises when we gauge the dual $U(1)_\phi$. The corresponding connection
\begin{equation}
  \widetilde{\mathscr A} = t_\phi ( l \dd \phi + \widetilde{l} \, \dd \widetilde\phi )
\end{equation}
leading to the dual generalised frame field
\begin{equation}
  \widetilde{\mathscr{E}}_L \equiv   \widetilde{E}_{\mathrm{L}1} + i   \widetilde{E}_{\mathrm{L}2} = \frac12 e^{i \widetilde{\phi}_-} \left( \partial_\theta - i \tan (\theta) \partial_{\widetilde{\phi}} + k \dd \theta - i k \cot (\theta) \dd \widetilde{\phi} \right) \,,
\end{equation}
where the specific combination is obtained from the latter after replacing\footnote{The flipping of sign in $k$ can be seen as e.g. the substitution $l \rightarrow - l$  while keeping $\widetilde{l}$ fixed. Together with the flipping in $\widetilde{\phi}$, this amounts to $\widetilde{\phi}_- \rightarrow - \widetilde{\phi}_+$.} $\widetilde{\phi} \rightarrow - \widetilde{\phi}$ and $k \rightarrow -k$, with $\widetilde{\phi}_\pm = ( 1 \pm \widetilde{l}/l ) \widetilde{\phi}$ and the generalised connection $\Omega = t_\phi ( k_{\widetilde{\phi}} - \widetilde{l} \dd \widetilde{\phi} )$. Again the connection $\mathscr{A}$ was chosen such that this frame field reproduces the monodromy in eq.~\eqref{eqn:monodromy}. More specifically, we obtain after restricting the domain of $\widetilde{\phi}$ to $0 \le \widetilde{\phi} < 2\pi / \widetilde{l}$ the monodromy
\begin{equation} 
  \mathscr{M}_A{}^B = \langle \widetilde{E}_A(\widetilde{\phi}=0) , \widetilde{E}^B(\widetilde{\phi}=2\pi/\widetilde{l}) \rangle =
  \left(e^{2 \pi ( t_{\widetilde{\phi}} - t_\phi )}\right)_A{}^B \,.
\end{equation}
Contracting $\widetilde{E}_A$ with the generalised metric gives rise to the T-dual metric and dilaton
\begin{equation}
  \dd s^2 = k ( \dd \theta^2 + \cot^2 \theta \dd \widetilde\phi^2 ) \,, \qquad
  e^{\Phi} = \frac{g_\mathrm{s}}{\sqrt{k}} \frac1{\sin\theta} \,,
\end{equation}
when $\mathcal{H}_{AB} = 2/k \, \delta_{AB}$ or the corresponding metric for its $\lambda$-deformation when $\mathcal{H}_{AB}$ is chosen as in eq.~\eqref{eq:Hforlambda}.

Changing $\theta$ to $\theta' = \pi/2 - \theta$ results in the same form of the metric and dilaton as in \eqref{eqn:metricdilatonparaferm}. But now the boundary of the disk and its center get exchanged. Similarly now the $U(1)_\phi$ monodromy generates a non-vanishing $\widetilde{\mathscr{F}}$ in the center,
\begin{equation}
  e^{\oint_{D^2} \widetilde{\mathscr{F}}} = e^{-2 \pi t_\phi}\,.
\end{equation}
 
Because $t_{\widetilde{\phi}}$ and $t_\phi$ are related by an $O(1,1)$ transformation the generalised geometries which are described by $E_A$ and $\widetilde{E}_A$ on a disk have identical generalised curvature and torsion. By construction they also share the same monodromy. In fact they correspond to the $\mathbb{Z}_l$ and $\mathbb{Z}_{\widetilde{l}}$ orbifold of the level $k$ parafermion CFT and its T-dual.

For $SL(2,\mathbb{R})/U(1)$ the derivation of the generalised frame fields proceeds along the same lines and for brevity we will not repeat it here. 

\subsection{NS5-branes on a circle}
A non-vanishing field strength $\mathscr{F}$, like the one we encountered in the last subsection, can result in a source for the $H$-flux. However this is only possible if the dimension of the target space is greater than three. Hence, in the following the $SL(2,\mathbb{R})/U(1)$ and the $SU(2)/U(1)$ geometries are combined along the line of \cite{Sfetsos:1998xd} to obtain a target space which captures the near horizon geometry of $k$ NS5-branes equally distributed on a circle.
 
A possible starting point is the observation that the geometry we are looking for is T-dual to \cite{Sfetsos:1998xd}
\begin{equation}
  \begin{aligned}
    \dd s^2 &= k \left[ \dd \theta^2 + \cot^2 \theta (\dd\chi - \dd \phi)^2 + \dd   
      \rho^2 + \tanh^2 \rho \, \dd\chi^2 \right] \, , \\
    \dd s^2 &= k \left[ \dd \theta^2 + \tan^2 \theta \,\dd \omega^2 + \dd \rho^2 + 
      \coth^2 \rho ( \dd \omega + \dd \psi )^2 \right]\,,
  \end{aligned}
\end{equation}
with the identifications $\xi \sim \xi + 2 \pi / k$, $\phi \sim \phi + 2\pi$, $\psi \sim \psi + 2 \pi$ and $\omega \sim \omega + 2 \pi/k$, respectively. Remembering the approach from the last subsection we identify the monodromy around T-dual pair of variables in each T-duality frame. Hence, we find the monodromies  
\begin{equation}
  \mathscr{M}_\chi = \mathscr{M}_\psi = e^{ 2\pi (t_\chi - t_\psi)} 
    \qquad \text{and} \qquad
  \mathscr{M}_\phi = \mathscr{M}_\omega = e^{ 2\pi(t_\omega - t_\phi)}\,,
\end{equation}
where the generators $t_\phi$, $t_\psi$, $t_\chi$, $t_\omega$ correspond to the vector fields $k_\phi = \partial_\phi$, $k_\psi = \partial_\psi$, $k_\chi = k^{-1} \partial_\chi$ and $k_\omega = k^{-1} \partial_\omega$ for the parameterisation
\begin{equation}
  g = \bigl( e^{i \sigma_2 ( \psi + \omega + \chi )/2} e^{\sigma^3 \rho} e^{i \sigma_3 ( \psi + \omega - \chi )/2} , e^{i \sigma_3 ( \chi - \omega - \phi )/2} e^{i \sigma_1 \theta} e^{i \sigma_3 ( \phi + \omega - \chi )/2} \bigr)
\end{equation}
of $SL(2,\mathbb{R})\times SU(2)$ elements $g$. These monodromies imply the connection
\begin{equation}\label{eqn:ANS5}
  \mathscr{A} = t_\chi ( \dd \psi + k \,\dd \chi ) + t_\omega ( \dd \phi + k\, \dd\omega )\,,
\end{equation}
for which the generalised frame field is constructed analogously to the
previous subsection. To present the result in a compact form, we introduce the  combinations $x_\pm = \frac{\phi - (k \pm 1) \psi}{k}$ and $y_\pm =  \frac{\psi + (k  \pm 1) \phi}{k}$ of coordinates and write the frame field components in terms of the complex combinations
\begin{equation}
  \begin{aligned}
    \mathscr{E}_{L1} \equiv E_{L1} + i E_{L2} &= \frac12 e^{- i x_+} \left(k  \partial_\rho + i k \coth(\rho) \partial_\psi + \dd \rho + i \tanh(\rho) \dd \psi - i \coth(\rho) \dd \phi\right) \, , \\
\mathscr{E}_{L2} \equiv E_{L3}+ i E_{L4} &= \frac12 e^{- i y_-} \left(k \partial_\theta + i k \cot(\theta) \partial_\phi + \dd \theta + i \tan(\theta) \dd \psi - i \cot(\theta) \dd \phi\right) \, ,\\
  \end{aligned}
\end{equation}
The remaining combinations, namely $\mathscr{E}_{R1}$ and $\mathscr{E}_{R2}$, can be obtained from $\mathscr{E}_{L1}$ and $\mathscr{E}_{L2}$ respectively by performing the substitutions $k \rightarrow -k$ and $\psi \rightarrow - \psi$. Using again the canonical generalised metric $\mathcal{H}_{AB} = k/2 \delta_{AB}$, this generalised frame field gives rise to the metric, dilaton
\begin{equation}
  \dd s^2 = k \left[ \dd \rho^2 + \dd \theta^2 + \frac{\psi^2 \tanh^2 \rho \, \dd \phi^2 + \tan^2 \theta \, \dd \psi^2}{1+\tan^2{\theta}\tanh^2{\rho}} \right]\, ,
    \qquad
  e^\Phi = \frac{1}{\sqrt{\cosh^2 \rho - \sin^2 \theta}}
\end{equation}
and (local) $B$-field
\begin{equation}
  B = \frac{k\, \dd \psi \wedge \dd \phi}{1+\tan^2{\theta} \tanh^2{\rho}}\,.
\end{equation}
To see that this background is indeed connected to NS5-branes on a circle one can embed the solution into the transversal $\mathbb{R}^4$ with the coordinates
\begin{equation}\label{eqn:NS5R4}
  \begin{aligned}
    x^1 &= \rho_0 \cosh \rho \sin \theta \cos \psi \,, \qquad&
    x^2 &= \rho_0 \cosh \rho \sin \theta \sin \psi \,, \\
    x^3 &= \rho_0 \sinh \rho \cos \theta \cos \phi \,, &
    x^4 &= \rho_0 \sinh \rho \cos \theta \sin \phi\,. 
  \end{aligned}
\end{equation}
In these coordinates the metric, dilaton and $H$-flux reads
\begin{equation}
  \dd s^2 = h \sum_{i=1}^4 (\dd x^i)^2 \,, \quad
  e^{2 \Phi} = h \,, \quad \text{and} \quad
  H =  - \star \dd \log h\,,
\end{equation}
where $h$ denotes the harmonic function
\begin{equation}
  h = \frac{2 k}{\rho_0^2 ( \cos 2 \theta + \cosh 2 \rho )} \,.
\end{equation}
Magnetic source for the $H$-flux are localised where $\star \dd H = \Delta \log h$ does not vanish. This happens only when the denominator of $h$ vanishes, i.e. for $\rho=0$ and $\theta=\pi/2$.

The target space is a fibration of a $T^2$ over a strip parameterised by $0 \le \rho$ and $0\le\theta\le\pi/2$. The torus has two cycles $A$ ($0 \le \psi < 2\pi$, $\phi=$  const.) and $B$ ($0\le\phi< 2\pi$, $\psi=$ const.). It is readily checked that the integral of the connection in eq.~\eqref{eqn:ANS5} over these two cycles gives rise
\begin{equation}
  e^{\oint_A \mathscr{A}} = e^{2\pi t_\chi} \qquad \text{and} \qquad
  e^{\oint_B \mathscr{A}} = e^{2\pi t_\omega}\,.
\end{equation}
In order to obtain the corresponding fluxes, one has to find out where these cycles collapse. As it can be seen for the embedding \eqref{eqn:NS5R4} into $\mathbb{R}^4$, the $A$ cycle collapses for $\theta=0$ and the $B$ cycle for $\theta=\pi/2$ and $\rho=0$. By following the argumentation of the last subsection, this implies non-vanishing field strength on these submanifolds. More specifically, we find
\begin{equation}
  \int_0^\theta \oint_A \mathscr{F} = 2\pi t_\chi \,, \qquad
  \int_0^\rho \oint_B \mathscr{F} = 2\pi t_\omega \qquad \text{and} \qquad
  \int_\theta^{\pi/2} \oint_B \mathscr{F} = 2\pi t_\omega\,.
\end{equation}
Furthermore, the resulting field strength has to be compatible with the $H$-flux and satisfy
\begin{equation}
  \dd H = - c( \mathscr{F} \overset{\wedge}{,} \mathscr{F} )\,.
\end{equation}
For $\rho=0$ and $\theta=0$ there is no source for the $H$-flux. Hence, the combination $\mathscr{F}^\chi \wedge \mathscr{F}^\omega$ should not contribute to the RHS and we conclude that $c(t_\chi, t_\omega)$ has to vanish. Because the symmetric pairing $c$ has to be non-degenerate the two additional contributions $c(t_\chi, t_\chi)$ and $c(t_\omega, t_\omega)$ have to be non-zero. This implies that $\mathscr{F}^\chi \wedge \mathscr{F}^\chi$ vanishes and furthermore 
\begin{equation}
  c( t_\omega, t_\omega ) \int_{\mathbb{R}^4} \mathscr{F}^\omega \wedge \mathscr{F}^\omega = - 4 \pi^2 k\,.
\end{equation}

Again one might ask if there are other $H$-invariant choices for the generalised metric. The commutant subgroup of $H$ and $O(4,4)$ is $SO(4)$. However, this $SO(4)$ is a subgroup of the $SO(4)\times SO(4)$ subgroup which leaves the generalised metric invariant. Hence, in contrast to the last subsection the metric, $B$-field and dilaton of this generalised coset are rigid. The generators of the unbroken $SO(4)$ can be expressed in terms of six generalised complex structures which give rise to a generalised hyper K\"ahler structure.

\section{Conclusions and future directions}
Let us first briefly recap the key results. On $M=\widetilde{G}\backslash \mathdsl{D}$ we reviewed the construction of a set of generalised frame fields $E_A$ that realise the algebra $\mathfrak{d}$ of $\mathdsl{D}$ via the generalised Lie derivative.  Being rather more direct and index free, the presentation makes more transparent the results obtained previously in \cite{Demulder:2018lmj}. Armed with these we perform a reduction to $M=\widetilde{G}\backslash \mathdsl{D}/H$ in which we quotient by a second isotropic subgroup $H$ provided suitable generalised reductiveness conditions are imposed. In this setting we construct both a generalised frame $E_A$ and a compensating $H$-valued generalised connection $\Omega^\alpha$ which is used to define a spin-connection piece of a connection $\nabla$ obeying the vielbein postulate $\nabla E_A =0$. The generalised torsion of $\nabla$, in flat indices, is determined entirely by a selection of the structure constants of $\mathfrak{d}$. An important intermediate step in the derivation of $E_A$ and $\Omega$ is the introduction of an $H$-valued flat connection $\mathscr{A}$ on $\widetilde{G}\backslash \mathdsl{D}$ whose pullback to $M=\widetilde{G}\backslash \mathdsl{D}/H$ locally vanishes. In general however it is not possible to find such a flat connection globally. As we illustrate with two examples, the resulting field strength $\mathscr{F}$ gives rise to additional, localised vector multiplets and is related to the global properties of the generalised frame field. Moreover these can lead to NS5-sources manifested in a violation of the Bianchi identity for the NS three-form $H$. We also encounter the situation where the action of $H$ has fixed points and the geometry exhibits singularities.

To give the above results some context we evaluate examples related to gauged-WZW models. For parafermionic theories and their orbifolds we provide generalised frame fields (which have quite simple expressions) for both vector and axial $U(1)$ gaugings.  We show that the generalised metric admits deformations and indeed that these can be identified with the integrable $\lambda$-deformations introduced in \cite{Sfetsos:2013wia}.  An elaboration of this is to consider the NS5-branes on a circle realised as a gauged-WZW model in \cite{Sfetsos:1998xd}.  Here we find equally elegant generalised frame fields describing a generalised hyper-K\"ahler manifold. Unlike the parafermonic example we find this geometry to be rigid i.e. we show it does not appear to admit a $\lambda$-type deformation.

This work opens up many interesting avenues for exploration both in terms of the formal aspects and their applications. Here we have only glimpsed at the enhancement of the generalised tangent bundle needed to accommodate fixed points and singularities in the geometry. A key goal is to make more explicit the linkage to anomaly cancelation via localised sources. Related to this, one hopes to understand in this framework the role that instantons are known to play in correcting T-duals in which the $U(1)$ actions degenerate and allow for string unwinding \cite{Gregory:1997te,Hori:1999me,Tong:2002rq,Harvey:2005ab}. Likely crucial to this will be the holonomies of the connection $\mathscr{A}$ which can depend on both original and  T-dual variables, and in this work we have made this somewhat explicit for the case of Abelian subgroups. We expect the cases with  $H$ non-Abelian to be even richer and more subtle.  

In terms of applications, these results should extend naturally to the RR sector of 10-dimensional supergravity and have utility in holography. It has been well established that non-Abelian T-duality can have applications in constructing new examples of holographic solutions (for a recent review and further references see  \cite{Thompson2019}).  With the tools presented here, upgraded to a full type II supergravity situation,  it can become viable to search for new interesting backgrounds obtained as Poisson-Lie duals on cosets.

For the case of $M=\widetilde{G}\backslash \mathdsl{D}$, allowing the generalised metric ${\cal H}_{AB}$ to depend on external coordinates specifies a consistent truncation Ansatz resulting in a maximal lower dimensional gauged supergravity.  Here for  $M=\widetilde{G}\backslash \mathdsl{D}/H$ one expects that ${\cal H}_{AB}$ should depend on external coordinates and also that vector multiplets will be induced by allowing the localised gauge fields $\mathscr{A}$ to have dynamics; the lower dimensional supergravity will not be maximal.  Making this precise, and understanding which lower dimensional supergravity theories can be obtained with this construction will be important. As a final note, let us conclude that here our attention has been restricted to the generalised tangent bundle $TM \oplus T^\star M$; one should anticipate that similar constructions will prove beneficial in the M-theoretic exceptional generalised geometry context.

\section*{Acknowledgements}
We like to thank David Berman, Charles Strickland-Constable and Daniel Waldram for comments on the draft. FH and DCT would like to express a special thanks to the Mainz Institute of Theoretical Physics where they were discussing this project during the program ``Holography, Generalised Geometry and Duality''. FH acknowledge the hospitality of the Galileo Galilei Institute during the workshop ``Integrable Effective Field Theories and Their Holographic Description''. DCT is supported by a Royal Society University Research Fellowship {\em Generalised Dualities in String Theory and Holography} URF 150185 and in part by STFC grant ST/P00055X/1 and in part by the ``FWO-Vlaanderen'' through the project G006119N and by the Vrije Universiteit Brussel through the Strategic Research Program ``High-Energy Physics''.  SD acknowledges the FWO aspirant fellowship and Max-Planck-Society for support. GP is supported by a Royal Society  Enhancement Award RGF/EA/180176. 

\bibliography{literaturNew}

\providecommand{\href}[2]{#2}\begingroup\raggedright\begin{thebibliography}{10}

\bibitem{Hitchin:2004ut}
N.~Hitchin, {\it {Generalized Calabi-Yau manifolds}},  {\em Quart. J. Math.}
  {\bf 54} (2003) 281--308, [\href{http://arxiv.org/abs/math/0209099}{{\tt
  math/0209099}}].

\bibitem{GualtieriThesis}
M.~{Gualtieri}, {\it {Generalized complex geometry}},
  \href{http://arxiv.org/abs/math/0401221}{{\tt math/0401221}}.

\bibitem{Hull:2007zu}
C.~M. Hull, {\it {Generalised Geometry for M-Theory}},  {\em JHEP} {\bf 07}
  (2007) 079, [\href{http://arxiv.org/abs/hep-th/0701203}{{\tt
  hep-th/0701203}}].

\bibitem{Pacheco:2008ps}
P.~Pires~Pacheco and D.~Waldram, {\it {M-theory, exceptional generalised
  geometry and superpotentials}},  {\em JHEP} {\bf 09} (2008) 123,
  [\href{http://arxiv.org/abs/0804.1362}{{\tt arXiv:0804.1362}}].

\bibitem{Coimbra:2012af}
A.~Coimbra, C.~Strickland-Constable, and D.~Waldram, {\it {Supergravity as
  Generalised Geometry II: $E_{d(d)} \times \mathbb{R}^+$ and M theory}},  {\em
  JHEP} {\bf 03} (2014) 019, [\href{http://arxiv.org/abs/1212.1586}{{\tt
  arXiv:1212.1586}}].

\bibitem{Hohm:2010pp}
O.~Hohm, C.~Hull, and B.~Zwiebach, {\it {Generalized metric formulation of
  double field theory}},  {\em JHEP} {\bf 08} (2010) 008,
  [\href{http://arxiv.org/abs/1006.4823}{{\tt arXiv:1006.4823}}].

\bibitem{Berman:2010is}
D.~S. Berman and M.~J. Perry, {\it {Generalized Geometry and M theory}},  {\em
  JHEP} {\bf 06} (2011) 074, [\href{http://arxiv.org/abs/1008.1763}{{\tt
  arXiv:1008.1763}}].

\bibitem{Hohm:2013pua}
O.~Hohm and H.~Samtleben, {\it {Exceptional Form of D=11 Supergravity}},  {\em
  Phys. Rev. Lett.} {\bf 111} (2013) 231601,
  [\href{http://arxiv.org/abs/1308.1673}{{\tt arXiv:1308.1673}}].

\bibitem{Lee:2014mla}
K.~Lee, C.~Strickland-Constable, and D.~Waldram, {\it {Spheres, generalised
  parallelisability and consistent truncations}},  {\em Fortsch. Phys.} {\bf
  65} (2017), no.~10-11 1700048, [\href{http://arxiv.org/abs/1401.3360}{{\tt
  arXiv:1401.3360}}].

\bibitem{Hohm:2014qga}
O.~Hohm and H.~Samtleben, {\it {Consistent Kaluza-Klein Truncations via
  Exceptional Field Theory}},  {\em JHEP} {\bf 01} (2015) 131,
  [\href{http://arxiv.org/abs/1410.8145}{{\tt arXiv:1410.8145}}].

\bibitem{Cassani:2019vcl}
D.~Cassani, G.~Josse, M.~Petrini, and D.~Waldram, {\it {Systematics of
  consistent truncations from generalised geometry}},
  \href{http://arxiv.org/abs/1907.06730}{{\tt arXiv:1907.06730}}.

\bibitem{Scherk:1978ta}
J.~Scherk and J.~H. Schwarz, {\it {Spontaneous Breaking of Supersymmetry
  Through Dimensional Reduction}},  {\em Phys. Lett.} {\bf 82B} (1979) 60--64.

\bibitem{Scherk:1979zr}
J.~Scherk and J.~H. Schwarz, {\it {How to Get Masses from Extra Dimensions}},
  {\em Nucl. Phys.} {\bf B153} (1979) 61--88.

\bibitem{Grana:2012rr}
M.~Grana and D.~Marques, {\it {Gauged Double Field Theory}},  {\em JHEP} {\bf
  04} (2012) 020, [\href{http://arxiv.org/abs/1201.2924}{{\tt
  arXiv:1201.2924}}].

\bibitem{Geissbuhler:2011mx}
D.~Geissbuhler, {\it {Double Field Theory and N=4 Gauged Supergravity}},  {\em
  JHEP} {\bf 11} (2011) 116, [\href{http://arxiv.org/abs/1109.4280}{{\tt
  arXiv:1109.4280}}].

\bibitem{Aldazabal:2011nj}
G.~Aldazabal, W.~Baron, D.~Marques, and C.~Nunez, {\it {The effective action of
  Double Field Theory}},  {\em JHEP} {\bf 11} (2011) 052,
  [\href{http://arxiv.org/abs/1109.0290}{{\tt arXiv:1109.0290}}]. [Erratum:
  JHEP11,109(2011)].

\bibitem{Grana:2008yw}
M.~Grana, R.~Minasian, M.~Petrini, and D.~Waldram, {\it {T-duality, Generalized
  Geometry and Non-Geometric Backgrounds}},  {\em JHEP} {\bf 04} (2009) 075,
  [\href{http://arxiv.org/abs/0807.4527}{{\tt arXiv:0807.4527}}].

\bibitem{Dibitetto:2012rk}
G.~Dibitetto, J.~J. Fernandez-Melgarejo, D.~Marques, and D.~Roest, {\it
  {Duality orbits of non-geometric fluxes}},  {\em Fortsch. Phys.} {\bf 60}
  (2012) 1123--1149, [\href{http://arxiv.org/abs/1203.6562}{{\tt
  arXiv:1203.6562}}].

\bibitem{Demulder:2018lmj}
S.~Demulder, F.~Hassler, and D.~C. Thompson, {\it {Doubled aspects of
  generalised dualities and integrable deformations}},  {\em JHEP} {\bf 02}
  (2019) 189, [\href{http://arxiv.org/abs/1810.11446}{{\tt arXiv:1810.11446}}].

\bibitem{Hassler:2019wvn}
F.~Hassler, D.~L{\"u}st, and F.~J. Rudolph, {\it {Para-Hermitian Geometries for
  Poisson-Lie Symmetric $\sigma$-models}},  {\em JHEP} {\bf 10} (2019) 160,
  [\href{http://arxiv.org/abs/1905.03791}{{\tt arXiv:1905.03791}}].

\bibitem{Klimcik:1995ux}
C.~Klimcik and P.~Severa, {\it {Dual nonAbelian duality and the Drinfeld
  double}},  {\em Phys. Lett.} {\bf B351} (1995) 455--462,
  [\href{http://arxiv.org/abs/hep-th/9502122}{{\tt hep-th/9502122}}].

\bibitem{Klimcik:1995dy}
C.~Klimcik and P.~Severa, {\it {Poisson-Lie T duality and loop groups of
  Drinfeld doubles}},  {\em Phys. Lett.} {\bf B372} (1996) 65--71,
  [\href{http://arxiv.org/abs/hep-th/9512040}{{\tt hep-th/9512040}}].

\bibitem{Hassler:2017yza}
F.~Hassler, {\it {Poisson-Lie T-Duality in Double Field Theory}},
  \href{http://arxiv.org/abs/1707.08624}{{\tt arXiv:1707.08624}}.

\bibitem{Klimcik:1996np}
C.~Klimcik and P.~Severa, {\it {Dressing cosets}},  {\em Phys. Lett.} {\bf
  B381} (1996) 56--61, [\href{http://arxiv.org/abs/hep-th/9602162}{{\tt
  hep-th/9602162}}].

\bibitem{Baraglia:2013wua}
D.~Baraglia and P.~Hekmati, {\it {Transitive Courant Algebroids, String
  Structures and T-duality}},  {\em Adv. Theor. Math. Phys.} {\bf 19} (2015)
  613--672, [\href{http://arxiv.org/abs/1308.5159}{{\tt arXiv:1308.5159}}].

\bibitem{Coimbra:2014qaa}
A.~Coimbra, R.~Minasian, H.~Triendl, and D.~Waldram, {\it {Generalised geometry
  for string corrections}},  {\em JHEP} {\bf 11} (2014) 160,
  [\href{http://arxiv.org/abs/1407.7542}{{\tt arXiv:1407.7542}}].

\bibitem{Aldazabal:2015yna}
G.~Aldazabal, M.~Graña, S.~Iguri, M.~Mayo, C.~Nuñez, and J.~A. Rosabal, {\it
  {Enhanced gauge symmetry and winding modes in Double Field Theory}},  {\em
  JHEP} {\bf 03} (2016) 093, [\href{http://arxiv.org/abs/1510.07644}{{\tt
  arXiv:1510.07644}}].

\bibitem{Aldazabal:2017wbk}
G.~Aldazabal, E.~Andres, M.~Mayo, and J.~A. Rosabal, {\it {Gauge symmetry
  enhancing-breaking from a Double Field Theory perspective}},  {\em JHEP} {\bf
  07} (2017) 045, [\href{http://arxiv.org/abs/1704.04427}{{\tt
  arXiv:1704.04427}}].

\bibitem{Blair:2018lbh}
C.~D.~A. Blair, E.~Malek, and D.~C. Thompson, {\it {O-folds: Orientifolds and
  Orbifolds in Exceptional Field Theory}},  {\em JHEP} {\bf 09} (2018) 157,
  [\href{http://arxiv.org/abs/1805.04524}{{\tt arXiv:1805.04524}}].

\bibitem{Sfetsos:1998xd}
K.~Sfetsos, {\it {Branes for Higgs phases and exact conformal field theories}},
   {\em JHEP} {\bf 01} (1999) 015,
  [\href{http://arxiv.org/abs/hep-th/9811167}{{\tt hep-th/9811167}}].

\bibitem{Severa:2018pag}
P.~Severa and F.~Valach, {\it {Courant algebroids, Poisson-Lie T-duality, and
  type II supergravities}},  \href{http://arxiv.org/abs/1810.07763}{{\tt
  arXiv:1810.07763}}.

\bibitem{bursztyn2007reduction}
H.~Bursztyn, G.~R. Cavalcanti, and M.~Gualtieri, {\it {Reduction of Courant
  algebroids and generalized complex structures}},  {\em Adv. Math.} {\bf 211}
  (2007) 726--765, [\href{http://arxiv.org/abs/math/0509640}{{\tt
  math/0509640}}].

\bibitem{figueroa1994equivariant}
J.~M. Figueroa-O'Farrill and S.~Stanciu, {\it Equivariant cohomology and gauged
  bosonic sigma-models},  {\em arXiv preprint hep-th/9407149} (1994).

\bibitem{hull1989gauged}
C.~Hull and B.~Spence, {\it The gauged nonlinear sigma model with wess-zumino
  term},  {\em Physics Letters B} {\bf 232} (1989), no.~2 204--210.

\bibitem{de1987new}
B.~de~Wit, C.~Hull, and M.~Ro{\v{c}}ek, {\it New topological terms in gauge
  invariant actions},  {\em Physics Letters B} {\bf 184} (1987), no.~2-3
  233--238.

\bibitem{Hohm_2013}
O.~Hohm and B.~Zwiebach, {\it Towards an invariant geometry of double field
  theory},  {\em Journal of Mathematical Physics} {\bf 54} (Mar, 2013) 032303.

\bibitem{Geissbuhler:2013uka}
D.~Geissbuhler, D.~Marques, C.~Nunez, and V.~Penas, {\it {Exploring Double
  Field Theory}},  {\em JHEP} {\bf 06} (2013) 101,
  [\href{http://arxiv.org/abs/1304.1472}{{\tt arXiv:1304.1472}}].

\bibitem{Hohm:2011ex}
O.~Hohm and S.~K. Kwak, {\it {Double Field Theory Formulation of Heterotic
  Strings}},  {\em JHEP} {\bf 06} (2011) 096,
  [\href{http://arxiv.org/abs/1103.2136}{{\tt arXiv:1103.2136}}].

\bibitem{Bardacki:1990wj}
K.~Bardakci, M.~J. Crescimanno, and E.~Rabinovici, {\it {Parafermions From
  Coset Models}},  {\em Nucl. Phys.} {\bf B344} (1990) 344--370.

\bibitem{Fateev:1985mm}
V.~A. Fateev and A.~B. Zamolodchikov, {\it {Parafermionic Currents in the
  Two-Dimensional Conformal Quantum Field Theory and Selfdual Critical Points
  in Z(n) Invariant Statistical Systems}},  {\em Sov. Phys. JETP} {\bf 62}
  (1985) 215--225. [Zh. Eksp. Teor. Fiz.89,380(1985)].

\bibitem{SfetsosSiamposThompson}
K.~Sfetsos, K.~Siampos, and D.~C. Thompson, {\it {Generalised integrable
  $\lambda$- and $\eta$-deformations and their relation}},  {\em Nucl. Phys.}
  {\bf B899} (2015) 489--512, [\href{http://arxiv.org/abs/1506.05784}{{\tt
  arXiv:1506.05784}}].

\bibitem{Sfetsos:2013wia}
K.~Sfetsos, {\it {Integrable interpolations: From exact CFTs to non-Abelian
  T-duals}},  {\em Nucl. Phys.} {\bf B880} (2014) 225--246,
  [\href{http://arxiv.org/abs/1312.4560}{{\tt arXiv:1312.4560}}].

\bibitem{Driezen:2019ykp}
S.~Driezen, A.~Sevrin, and D.~C. Thompson, {\it {Integrable asymmetric
  $\lambda$-deformations}},  {\em JHEP} {\bf 04} (2019) 094,
  [\href{http://arxiv.org/abs/1902.04142}{{\tt arXiv:1902.04142}}].

\bibitem{Gregory:1997te}
R.~Gregory, J.~A. Harvey, and G.~W. Moore, {\it {Unwinding strings and t
  duality of Kaluza-Klein and h monopoles}},  {\em Adv. Theor. Math. Phys.}
  {\bf 1} (1997) 283--297, [\href{http://arxiv.org/abs/hep-th/9708086}{{\tt
  hep-th/9708086}}].

\bibitem{Hori:1999me}
K.~Hori, {\it {D-branes, T duality, and index theory}},  {\em Adv. Theor. Math.
  Phys.} {\bf 3} (1999) 281--342,
  [\href{http://arxiv.org/abs/hep-th/9902102}{{\tt hep-th/9902102}}].

\bibitem{Tong:2002rq}
D.~Tong, {\it {NS5-branes, T duality and world sheet instantons}},  {\em JHEP}
  {\bf 07} (2002) 013, [\href{http://arxiv.org/abs/hep-th/0204186}{{\tt
  hep-th/0204186}}].

\bibitem{Harvey:2005ab}
J.~A. Harvey and S.~Jensen, {\it {Worldsheet instanton corrections to the
  Kaluza-Klein monopole}},  {\em JHEP} {\bf 10} (2005) 028,
  [\href{http://arxiv.org/abs/hep-th/0507204}{{\tt hep-th/0507204}}].

\bibitem{Thompson2019}
D.~C. Thompson, {\it {An Introduction to Generalised Dualities and their
  Applications to Holography and Integrability}},  in {\em {18th Hellenic
  School and Workshops on Elementary Particle Physics and Gravity (CORFU2018)
  Corfu, Corfu, Greece, August 31-September 28, 2018}}, 2019.
\newblock \href{http://arxiv.org/abs/1904.11561}{{\tt arXiv:1904.11561}}.

\end{thebibliography}\endgroup
   
\bibliographystyle{JHEP}
\end{document}